\documentclass[11pt]{article}

\usepackage{latexsym,cite,nicefrac}
\usepackage[dvips]{graphicx}
\usepackage{color}
\usepackage[affil-it]{authblk}
\usepackage[cp1252]{inputenc}

\newtheorem{theorem}{Theorem}[section]
\newtheorem{corollary}{Corollary}[section]
\newtheorem{lemma}{Lemma}[section]
\newtheorem{proposition}{Proposition}[section]

\begin{document}

\title{A general approach to deriving diagnosability results of interconnection networks\thanks{Preliminary versions of some results of this paper were announced (without proofs) at 2019 International Conference on Modeling, Simulation, Optimization and Algorithm\ \cite{CMQS2019}.}}

\author[1]{Eddie Cheng}
\author[2]{Yaping Mao}
\author[3]{Ke Qiu}
\author[4]{Zhizhang Shen\thanks{Corresponding author\ E-mail: zshen\@plymouth.edu}}

\affil[1]{%
Department of Mathematics and Statistics, Oakland University, Rochester, MI 48309, USA}
\affil[2]{%
School of Mathematics and Statistics, Qinghai Normal University, Xining, Qinghai, 810001, CHINA}
\affil[3]{%
Department of Computer Science, Brock University, St. Catharines, Ontario, L2S 3A1, CANADA}
\affil[4]{%
Department of Computer Science and Technology, Plymouth State University, Plymouth, NH 03264, USA}

\maketitle

\begin{abstract}
We generalize an approach to deriving  diagnosability results of various interconnection networks in terms of the popular $g$-good-neighbor and $g$-extra fault-tolerant models, as well as mainstream diagnostic models such as the  PMC and the MM* models. 

As demonstrative examples, we show how to follow this constructive, and effective, process to derive the $g$-extra diagnosabilities of the hypercube, the $(n, k)$-star, and the arrangement graph. These results agree with those achieved individually, without duplicating structure independent technical details. Some of them come with a larger applicable range than those already known, and the result for the arrangement graph in terms of the MM* model is new. 
\bigskip

This is an Accepted Manuscript of an article published by Taylor \& Francis Group in the {\em International Journal of Parallel, Emergent \& Distributed Systems} on 03/29/2022, available online: \verb+https://doi.org/10.1080/17445760.2022.2060977+.
\end{abstract}

\section*{Keywords}
Fault tolerance; diagnosability; $g$-good-neighbor diagnosability; $g$-extra diagnosability; the hypercube graph; the $(n,k)$-star graph; the arrangement graph

\section{Introduction}

A rapid and consistent technical progress in computing technology has made  multi-processor systems a reality, where an inter-processor communication enabling network structure  plays a central role. It is unavoidable that some of the processing nodes in such a system become faulty, potentially disabling its integrity. To cope with such a situation, it is mandatory to develop technology to identify, and then correct/replace, faulty nodes in these network systems in order to restore its normal operation. For obvious reasons, one would want to have a self-diagnosable system where the computing nodes are able to detect faulty ones themselves. The maximum number of faulty nodes that can be so identified in an interconnection system is called its {\em diagnosability.} An ideal system should come with a large diagnosability, indicating a greater fault tolerance in the sense that, even this many processing nodes fail, they can still be identified and corrected so that the normal functionality of the system can be restored. The process of deriving such diagnosability results helps us to choose a fault-tolerant, thus sustainable, interconnection networks to meet our daily needs. It hence has attracted much attention in the research community over an extended period of time. 

Naturally, the diagnosability of a network depends on its topological structure, the intended fault-tolerant model, and the diagnostic model deemed appropriate. The topology of an interconnection network system is usually modeled with a connected graph $G(V, E),$ where $V,$ the set of {\em vertices of} $G,$ represents a collection of processing nodes; and $E,$ the set of {\em edges,} the connection between pairs of nodes in such a system. Many network topologies have been suggested and studied, including such influential graph structures as the hypercube\ \cite{Harary1988} and many of its variants,  the star graph\ \cite{Akers1989},  the bubble-sort graph\ \cite{Chou1996}, the arrangement graph\ \cite{Day1992}, and the $(n, k)$-star graph\ \cite{Chiang1995}.

By a {\em neighbor of a vertex,} $v,$ {\em in a graph} $G,$ we mean a vertex $u$ such that $(u, v)$ is an edge in $G$; and, the {\em degree of a vertex,} $v,$ is simply the number of its neighbors in $G.$ We also say $G$ is {\em $r$-regular} if the degrees of all its vertices equal $r.$   Clearly, there is no way to tell if a vertex is faulty, if all its neighbors are. Such a vertex then cannot be used to judge the faulty status of any of its neighbors.  The diagnosability of a graph $G$ is thus no more than $\delta(G),$ the minimum degree of any vertex in $G$\ \cite{Hakimi1974,Preparata1967,Lin2008}, which is, unfortunately, neither theoretically interesting nor practically satisfactory.  On the other hand, since this scenario that all the neighbors of every node\ (vertex) could be faulty, as implied by an {\em unrestrictive fault-tolerant model,} is highly unlikely,  several more sophisticated and realistic fault-tolerant models have since been suggested. Let a  {\em faulty set} be a collection of vertices $F \subset V,$ which are effectively removed from $G.$ Such a faulty set is  {\em conditional faulty}\ \cite{Lai2005,CQS2012,Hong2012} when every vertex, faulty or not, has at least one {\em fault-free}  neighbor in the {\em survival graph} of $G-F.$ For $g \geq 0,$ a faulty set is  {\em $g$-good-neighbor faulty}\ \cite{Peng2012} when every fault-free vertex has at least $g$ fault-free neighbors in the survival graph; and,  a faulty set is  {\em $g$-extra faulty}\ \cite{Fabrega1996,Zhang2016} when every component in the survival graph contains at least $g+1$ vertices. 


There is also an ``edge version'' for both the $g$-good-neighbor, and the $g$-extra, faulty sets. For example,  a set of edges $F$ in a connected graph $G $ is called an {\em $g$-good-neighbor edge-cut} if the survival graph $G-F$ is disconnected and every fault-free vertex  has at least $g$ fault-free neighbors\ \cite{CHQS2019}\ \footnote{The original notion of a good-neighbor edge-cut of order $m$ as coined in\ \cite{CHQS2019} is that a set of edges $T$ in a connected graph $G$ is called a good-neighbor edge-cut of order $m$ if $G-F$ is disconnected and every vertex in $G-F$ has degree at least $m.$}. We will focus on the vertex version of these two notions in this paper. For various properties of the edge versions of various faulty sets, associated results, and their relationship with their vertex related cousins, readers are referred to\ \cite{CHQS2019}.

The often adopted  {\em comparison diagnostic model}, i.e., the MM* diagnostic model\ \cite{Sengupta1992}, places a restriction on the MM model\ \cite{Malek1980,Maeng1981}, so that every processing node, acting as a {\em testing node}, sends a test message to each and every pair of its distinct neighbors, referred to as the {\em tested nodes,} and then compares their responses. The fault status of the system can then be determined, and the faulty nodes identified, based on the comparison results so obtained. The PMC model\ \cite{Preparata1967} is another popular diagnostic model where every node sends a test message to each of its neighbors and obtain the final diagnostic information based on the received testing results. Various efficient algorithms to identify such  faulty sets have also been proposed in, e.g.\ \cite{Dahbura1984,Sullivan1984,Sengupta1992,Zhu2008}. With both the MM* and the PMC models, it is assumed that, when a testing node is faulty, responses from those tested nodes will be unreliable. On the other hand, the BMG model\ \cite{Barsi1976} assumes that, under such circumstances, the response from a tested node is always fault free, even if it is faulty; and uncertain when the tested node is fault free. 

A collection of all such test results obtained with a diagnostic model is called a {\em syndrome} of the diagnosis in $G.$  A subset $F\ (\subset V)$ is said to be {\em allowed for a syndrome}\ \cite{Dahbura1984},  or {\em compatible with a syndrome} \cite{Lai2005}, if it can be generated  when  all the vertices in $F$ are faulty and all those in $V \setminus F$  are fault free. Since faulty testing nodes lead to unreliable results, as observed in  \cite{Maeng1981,Malek1980}, two faulty sets may be compatible with the same syndrome, thus making such a faulty set unidentifiable, and the diagnostic process fallible. This observation leads to the notion of a graph being {\em $t$-diagnosable} \cite{Lai2005,Sengupta1992}: when up to $t$ faulty vertices in $G$ can be identified. And the {\em diagnosability of a graph} $G$, denoted by $t(G)$, is defined to be the maximum number of faulty vertices that $G$ can guarantee to identify in terms of this diagnostic model.

Because of their important role in both network theory and practice, many diagnosability results have appeared in literature.  Recent examples include $g$-extra diagnosability results for the hypercube\ \cite{Zhang2016}, the arrangement graph\ \cite{Xu2016}, the bubble-sort graph\ \cite{Wang2016b}, the $(n, k)$-star graph\ \cite{Lv2019}, and the hierarchical cubic network\ \cite{Liu2019}, all in terms of both PMC and/or MM* models. We notice that much of the {\em ad hoc} derivation details, as reported in those papers devoted to different structures are essentially shared among themselves, and even with the  results on the $g$-good-neighbor diagnosability derived for these structures\ \cite{Peng2012,Wang2016a,Wang2016d,Xu2017,CQS2019,Wang2019}.  While we have benefited greatly from studying many of these earlier results, we believe that it has reached a point that such a practice is no longer desirable. Indeed, such a ``wheel'' should be reused, but not reinvented every time.

In particular, we realize that many notions related to diagnosability are independent of both fault-tolerant, and diagnostic, models.  Thus, much of the reasoning behind the derivation of the diagnosability of a specific network structure under several fault-tolerant models, e.g., the $g$-good-neighbor and the $g$-extra models, are essentially the same for a specific diagnostic model such as the PMC or the MM* model. We also observe that, because of the relationships among various fault-tolerant models, results applicable to one  model might follow from existing ones pertinent to another. Furthermore, some of the recently developed proof techniques can also play a role in shortening  mechanical proofs of related diagnosability results. We thus believe it is time for us to generalize such common and mechanical parts, separate them from the structure dependent analysis, and investigate their applicability so that future research in this important and active  area, of a more creative nature, could focus on the important issues related to structure, fault tolerant, and diagnostic, models, but not on mundane derivation details. We do notice that several results of such a general and summarizing nature have already appeared in\ \cite{CQS2013,Cheng2018,Lin2018,CQS2019,Wang2019}. On the other hand, some of them carry a restriction of an existential nature, assuming the existence of a certain property, thus computationally expensive, and may not be effectively applicable. 

In this paper, we will continue an effort that we started in\ \cite{CQS2012,CQS2013}, focused on the conditional fault-tolerant model,  and in\ \cite{CQS2019},  focused on the $g$-good-neighbor fault-tolerant model, by following a constructive approach to explore, expose, and summarize such a general, commonly shared, and effectively applicable, diagnosability derivation process for both the $g$-good-neighbor and the $g$-extra models. We will also demonstrate the applicability of this process to derive several diagnosability results for various network structures in terms of the more recently suggested $g$-extra fault-tolerant model, under both the PMC and the MM* diagnostic models. 

The rest of this paper proceeds as follows: In Section~\ref{section:relation}, after presenting basic notions, we provide a general derivation to an existing result between various notions of diagnosability, and derive several related results, which were justified separately in\ \cite{Wang2016b}, to set the stage for Section~\ref{section:process}, where we summarize a general process of deriving diagnosability results shared by both the $g$-good-neighbor and the $g$-extra fault-tolerant models, in terms of either the PMC, or the MM*, model.  We demonstrate the value, and applicability, of this general process by deriving the $g$-extra diagnosability of the  hypercube graph  in Section~\ref{section:resultQ}, that of the   $(n, k)$-star graph in Section~\ref{section:resultNK},  and that of the arrangement graph in Section~\ref{section:resultA}. We conclude this paper  in Section~\ref{section:end}.

\section{Relationships among fault-tolerant models}
\label{section:relation}

Let $G(V, E)$ represent an interconnection network, and let $M$ stand for a certain fault-tolerant model, an $M$-faulty set of $G$ is a set, $F \subset V,$ consistent with $M.$ For example, $F (\subset V)$ is a $g$-extra faulty set if every component in $G-F$ contains at least $g+1$ vertices.  $G$ is called  {\em $M$ $t$-diagnosable in terms of a diagnostic model $D$, }if $G$ is diagnosable for each and every  $M$-faulty set of size at most $t$ in $D,$ where {\em $D$ refers to either the PMC model or the MM* model in the rest of this paper.} 
   
Let $F_1$ and $F_2$ be two distinct $M$-faulty sets, $F_1 \subset V(G)$ and $F_2 \subset V(G),$  the pair $(F_1, F_2)$ is {\em distinguishable} in $G$ if and only if they are not compatible with the same syndrome, thus identifiable.  They are {\em indistinguishable} if they are compatible with some syndrome. Then, as originally suggested in\ \cite{Preparata1967} and summarized later in\ \cite[Lemma~5]{Lai2005}, $t_M(G, D),$ the {\em $M$-diagnosability of $G,$ in terms of a diagnostic model $D,$} equals the maximum number $t$ such that, for all the distinct $M$-faulty set pairs $(F_1, F_2),$ such that $F_1 \subset V, F_2 \subset V,$ and $|F_1| \leq t, |F_2| \leq t,$   $(F_1, F_2)$ is distinguishable in terms of $D.$  

We notice that all the diagnosability notions related to  existing fault-tolerant models, including, unrestricted, conditional, $g$-good-neighbor, and $g$-extra, use this {\em maximum restriction.} We assume that all the $M$-diagnosability notions that we discuss in this paper also satisfy this requirement.

Thus,  the diagnosability problem, of determining this measurement of $t_M(G, D)$ for a given graph $G(V, E)$ in terms of a fault-tolerant model $M$ under a diagnostic model $D,$ really comes down to a decision problem in graph theory: Are two $M$-faulty sets of $V$ distinguishable under $D$? 

In this regard, the following  result specifies a necessary and sufficient condition of two faulty sets being distinguishable under the MM* model, where $F_1 \Delta F_2$ stands for $(F_1 \setminus F_2) \cup (F_2  \setminus F_1),$ i.e., the symmetric difference of $F_1$ and $F_2.$ We notice that this result has nothing to do with either the involved fault-tolerant model $M,$ or the size of such a faulty set. 
\begin{theorem}
\cite{Sengupta1992} 
\label{theorem:MMCase}
Let $G(V, E)$ be a graph, and let $F_1$ and $F_2$ be two distinct subsets  of $V,$ $F_1$ and $F_2$ are distinguishable under the MM* model if and only if at least one of the following three conditions is satisfied: 
\begin{itemize} 
  \item there are two distinct vertices $v$ and $w$ in $V \setminus (F_1 \cup F_2)$ and there is a vertex $x$ in $F_1 \Delta F_2$ such that $(v, w, x)$ is a path in $G;$  \item there are two distinct vertices $v$ and $x$ in $F_1 \setminus F_2$ and there is a vertex $w$ in $V \setminus (F_1 \cup F_2)$ such that $(v, w, x)$ is a path in $G;$ and 
  \item there are two distinct vertices $v$ and $x$ in $F_2 \setminus F_1$ and there is a vertex $w$ in $V \setminus (F_1 \cup F_2)$ such that $(v, w, x)$ is a path in $G.$
\end{itemize}
\end{theorem}

\begin{figure}[ht]
\begin{center}
\begin{picture}(80,60)(10,0)

\thicklines
  \put(0,0){\line(1,0){120}}
  \put(0,60){\line(1,0){120}}
  \put(0,0){\line(0,1){60}}
  \put(120,0){\line(0,1){60}}

\thinlines
  \put(50,30){\circle{40}} 
  \put(70,30){\circle{40}}

  \put(30,52){\makebox(0,0){\small{$F_1$}}}
  \put(90,52){\makebox(0,0){\small{$F_2$}}}

  \put(10,50){\makebox(0,0){\small{$w$}}}
  \put(45,40){\makebox(0,0){\small{$x$}}}
  \put(16,48){\line(3,-1){24}}
  \put(38,28){\makebox(0,0){\small{$v$}}}
  \put(16,48){\line(1,-1){18}}

 \put(95,6){\makebox(0,0){\small{$w$}}}
  \put(110,6){\makebox(0,0){\small{$v$}}}
  \put(92,12){\line(1,0){18}}
  \put(80,30){\makebox(0,0){\small{$x$}}}
  \put(92,12){\line(-1,1){12}}
 \end{picture}
 \end{center}
\caption{The MM* distinguishability}
 \label{figure:MM}
\end{figure}
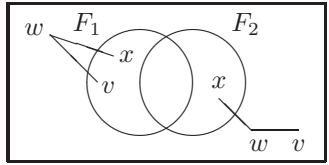
One of the two symmetric scenarios of Case~1, and Case~2, of Theorem~\ref{theorem:MMCase} are demonstrated in Figure~\ref{figure:MM}, while Case~3 is symmetric to Case~2.

The conditions associated with the PMC model  are somewhat simpler, as expected, since in the MM* model we test pairs of neighbors, but not just individual neighbors as in the PMC model. The involved situation is demonstrated in Figure~\ref{figure:PMC}.
\begin{theorem}
\cite{Dahbura1984}
\label{theorem:PMCCase}
Let $G(V, E)$ be a graph. For any two distinct subsets $F_1$ and $F_2$ of $V,$ $F_1$ and $F_2$ are distinguishable under the PMC model if and only if there exist a vertex $u$ in $V \setminus (F_1 \cup F_2)$ and another vertex $v$ in $F_1 \Delta F_2$ such that $(u, v)$ is an edge of $G.$ 
\end{theorem}
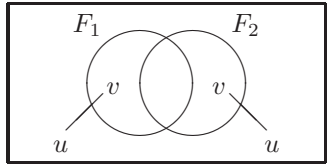
\begin{figure}[ht]
\begin{center}
\begin{picture}(80,60)(10,0)

\thicklines

  \put(0,0){\line(1,0){120}}
  \put(0,60){\line(1,0){120}}
  \put(0,0){\line(0,1){60}}
  \put(120,0){\line(0,1){60}}

\thinlines

  \put(50,30){\circle{40}} 
  \put(70,30){\circle{40}}

  \put(30,52){\makebox(0,0){\small{$F_1$}}}
  \put(90,52){\makebox(0,0){\small{$F_2$}}}

 \put(100,6){\makebox(0,0){\small{$u$}}}
 \put(80,28){\makebox(0,0){\small{$v$}}}
  \put(98,12){\line(-1,1){14}}

 \put(20,6){\makebox(0,0){\small{$u$}}}
 \put(40,28){\makebox(0,0){\small{$v$}}}
  \put(22,12){\line(1,1){14}}
 \end{picture}
\end{center}
 \label{figure:PMC}
 \caption{ The PMC distinguishability}
\end{figure}
To study the relationships between fault-tolerant models, we start with a general result, where the gist of the proof was used in\ \cite[Proposition~2.2]{Wang2016b} to derive several results of a similar nature, then apply this result to derive this collection of specific results to demonstrate the fact that the reasoning behind the same diagnostic issues among different fault-tolerant models are indeed shared. 
\begin{theorem}
\label{theorem:central}
Let $G$ be a connected graph, and let $t_{M_1}(G, D)$ and $t_{M_2}(G, D)$ be $M_1$-diagnosability, and $M_2$-diagnosability of $G$ in $D,$ respectively. If every $M_2$-faulty set in $G$ is also a $M_1$-faulty set in $G,$ then $t_{M_1}(G, D) \leq t_{M_2}(G, D).$ 
\end{theorem}

\noindent{\bf Proof:} Assume that $t_{M_1}(G, D)=t,$ and let $\mathcal{V}_1$\ (respectively, $\mathcal{V}_2$) be the collection of all the $M_1$-\ (respectively, $M_2$-) faulty sets $F$ in $G$ such that $|F| \leq t.$ Then $\mathcal{V}_2 \subseteq \mathcal{V}_1$ by assumption. Let $F_1$ and $F_2$ be two distinct sets, where $F_1, F_2 \in \mathcal{V}_2,$ then $F_1, F_2 \in \mathcal{V}_1.$ 

Since $|F_1|, |F_2| \leq t,$ and they are both $M_1$-faulty sets, by the assumed maximum restriction, $G$ is ${M_1}$ $t$-diagnosable in terms of $D.$ By definition, $(F_1, F_2)$ is distinguishable in $G$ in terms of the diagnostic model $D.$ Since both $F_1$ and $F_2$ are also $M_2$-faulty sets, by assumption of this result,  $G$ is also ${M_2}$ $t$-diagnosable in terms of $D.$ Finally, again, by the maximum restriction, $t_{M_2}(G, D)$ is the maximum value for which $G$ is $t_{M_2}$ $t$-diagnosable in $D.$ Hence $t_{M_2}(G, D) \geq t =t_{M_1}(G, D).$ 
\smallskip

The following result motivates most of the work along this line of diagnosability research since we naturally would like an interconnection network to be more fault-tolerant, i.e., to have a larger diagnosability. 
\begin{corollary}
\label{corollary:first}
\cite[Proposition~2.1]{Wang2016b}
Let $G$ be a system, and let $t(G, D)$ be the unrestricted diagnosability of $G,$ and let $t_M(G, D)$ be the $M$-diagnosability of $G,$ satisfying the maximum restriction, then, $t(G, D) \leq t_M(G, D).$
\end{corollary}

\noindent{\bf Proof:} Since any $M$-faulty set as associated with $t_M(G, D)$ is immediately an unrestricted faulty set, the result now follows Theorem~\ref{theorem:central}.  

\smallskip

The following two results, generalizing Corollary~\ref{corollary:first},  are intuitively true, since a stronger restriction often leads to a larger fault tolerance. Notice that with both the $0$-good-neighbor faulty sets and the 0-extra faulty sets, no restriction is placed on them, thus both 0-good-neighbor and 0-extra diagnosability reduce to the traditional unrestrictive diagnosability\ \cite[Section~4]{Peng2012}, i.e., the vertex connectivity of the involved graph. Hence, such more general notions, when $g \geq 1,$ do lead to a more generous characterization of the fault-tolerant properties of a graph. {\em We will assume $g \geq 1$ in the rest of this paper, unless explicitly pointed out otherwise. }
\begin{corollary}
\label{corollary:tg}
\cite[Proposition~2.1]{Wang2016b}
Let $G$ be a system, and let $t_g(G, D)$ be the associated $g$-good-neighbor diagnosability of $G$,  then $t_g(G, D) \leq t_{g'}(G, D), 0 \leq g \leq g'.$
\end{corollary}

\noindent{\bf Proof:} By definition, the $g$-good-neighbor diagnosability satisfies the maximum restriction. Moreover, if a fault-free vertex has at least $g'$ fault-free neighbors, it must have at least $g$ such neighbors when $0 \leq g \leq g'. $ Hence, a $g'$-good-neighbor faulty set is a $g$-good-neighbor faulty set, and the result follows Theorem~\ref{theorem:central}.

\begin{corollary}
\label{corollary:ttg}
\cite[Proposition~2.2]{Wang2016b}
Let $G$ be a system, and let $\overline{t}_g(G, D)$ be the $g$-extra diagnosability of $G,$ then $\overline{t}_g(G, D) \leq \overline{t}_{g'}(G, D), 0 \leq g \leq g'.$
\end{corollary}

\noindent{\bf Proof:} By definition, the $g$-extra diagnosability satisfies the maximum restriction. Moreover, if any component in $G-F$ contains at least $g'+1$ vertices, then such a component must contain at least $g+1$ vertices when $0 \leq g \leq g'.$ The result now follows Theorem~\ref{theorem:central}. 

\smallskip

The following two results show that the $g$-good-neighbor diagnosability of a graph is an upper bound of its $g$-extra diagnosability, except when $g=1.$ In particular, in Section~\ref{section:resultNK}, we will show how to make use of Corollary~\ref{corollary:gnE} to derive an upper bound of the $g$-extra diagnosability of the $(n, k)$-star graph with an existing $g$-good-neighbor diagnosability result for the same graph. 

\begin{corollary}
\label{corollary:gnE}
\cite[Theorem~2.2]{Wang2016b}
Let $G$ be a system, and let $g \geq 0,$ then $\overline{t}_g(G, D) \leq t_g(G, D).$
\end{corollary}

\noindent{\bf Proof:} Assume that $F$ is a $g$-good-neighbor faulty set in $G,$ then, every fault-free vertex has at least $g$ fault-free neighbors, thus, any component in $G-F$ must have at least $g+1$ fault-free vertices, i.e., $F$ is also a $g$-extra faulty set of $G.$ And the result follows from Theorem~\ref{theorem:central}. 
\begin{corollary}
\label{corollary:g1=tg1}
\cite[Theorem~2.3]{Wang2016b}
Let $G$ be a system,  then $\overline{t}_1(G, D)=t_{1}(G, D).$
\end{corollary}

\noindent{\bf Proof:} By Corollary~\ref{corollary:gnE}, $\overline{t}_1(G, D) \leq t_{1}(G, D).$ On the other hand, let $F$ be a 1-extra faulty set of $G,$ i.e., every component of $G-F$ contains at least two vertices, then each fault-free vertex in $G-F$ has at least one fault-free neighbor, thus, $F$ is also a 1-good-neighbor faulty set. The result now follows Theorem~\ref{theorem:central}. 

\smallskip

Notice that, for $g \geq 2,$ it does not generally hold that ${t}_g(G, D) \leq \overline{t}_g(G, D),$ since a $g$-extra faulty set is not necessarily a $g$-good-neighbor faulty set of $G.$ For example, if $F$ contains a path with $g+1$ vertices, it is certainly not a $g$-good-neighbor faulty set of $G.$  Indeed, the forthcoming Theorems~\ref{theorem:Ngd=2} and~\ref{theorem:2gtA} show that, for the arrangement graph, its 2-good-neighbor diagnosability is strictly larger than its 2-extra diagnosability. 

As we will discuss later in the paper, when deriving diagnosability, the case of $g=1$ is technically challenging as far as the MM* model is concerned, and often tedious. Examples include\ \cite[Lemma~5.2]{Wang2016b},\ \cite[Claim~A.1]{CQS2017b},\ \cite[Theorem~2]{Lin2018},\ \cite[Lemma~4.2]{CQS2019}, and\ \cite[Lemma~4.6]{Lv2019}. On the other hand, since the notion of $g$-good-neighbor fault-tolerant model was suggested earlier than that of $g$-extra diagnosability, more results for the former model have already been achieved, thus available when seeking related results in terms of the $g$-extra fault-tolerant model. Corollary~\ref{corollary:g1=tg1} becomes valuable in this regard, and Theorems~\ref{theorem:tQMM}, Corollary~\ref{corollary:tgN1}, and Theorem~\ref{theorem:tgA1MM} provide  examples of its application. 

It is important to point out that the notion of the $1$-good-neighbor conditional diagnosability is not a straightforward generalization of the conditional diagnosability. Indeed, the notion of the $1$-good-neighbor conditional diagnosability is less restrictive in the sense that the related 1-good-neighbor faulty set only requires that a non-faulty vertex have at least one non-faulty neighbor, while a conditional faulty set requires that any vertex, faulty or not, have at least one non-faulty neighbor. Thus, a conditional faulty set is immediately a 1-good-neighbor faulty set, but the other direction is not necessarily true. As a result, the 1-good-neighbor diagnosability of a graph is a lower bound of its conditional diagnosability, which also naturally follows from the general result, i.e., Theorem~\ref{theorem:central}.
\begin{corollary}
\cite{Gu2016,Wang2016b}
Let $G$ be a system, and let $t_c(G, D)$ be its conditional diagnosability, then $t_1(G, D) \leq t_c(G, D).$
\end{corollary}

\noindent{\bf Proof:} Every conditional faulty set of a system $G$ is a 1-good-neighbor faulty set of $G,$ since if every vertex in $V(G)$ has a fault-free neighbor, every fault-free vertex in $G-F$ has at least one fault-free neighbor. The result now follows from Theorem~\ref{theorem:central}. 
\smallbreak

For example, for $n \geq 4, k \in [3, n),$ the 1-good-neighbor diagnosability of the $(n, k)$-star graph is $n+k-2$\ \cite[Theorem~5.3]{CQS2019}, while its conditional diagnosability is $n+2k-5$\cite[Corollary~4.1]{CQS2013}.

\smallskip

By Corollary~\ref{corollary:g1=tg1}, we immediately have the following result. 
\begin{corollary}
Let $G$ be a system, $\overline{t}_1(G, D) \leq t_c(G, D).$
\end{corollary}

We would like to point out that, in establishing many of the results later in this paper, we require that the fault-tolerant model be in the format that {\em only a fault-free vertex has a property in the survival graph.} We notice that this  {\em ``fault-free'' requirement} is consistent with both the $g$-good-neighbor and the $g$-extra fault-tolerant models, although not with the conditional fault-tolerant model as pointed out earlier.  This is certainly not a surprise as the concept of conditional fault-tolerant model predates both the $g$-good-neighbor and $g$-extra models. 

\section{A general process of deriving diagnosability results}
\label{section:process}

Recall that an {\em $M$-faulty set} is just a faulty vertex set $F$ in a graph $G(V, E),$ related to a certain fault-tolerant model $M.$  An $M$-faulty set $F$ is also an {\em $M$-cut} if $G-F$ is disconnected. For example, a $g$-extra faulty set, $F,$ is also a $g$-extra cut if $G-F$ is disconnected, where every connected component contains at least $g+1$ vertices. Although an $M$-faulty set does not need to be an $M$-faulty cut, the construction of such an $M$-faulty cut turns out to be a crucial step to derive an $M$-diagnosability result, especially its upper bound. 

The size of a minimum $M$-faulty cut of a graph $G,$ on the other hand, is referred to as its {\em $M$-connectivity,}  denoted by $\kappa_M(G)$. The $M$-connectivity of a graph depends on its topology, and is often tedious and challenging to derive, but it plays a critical role in deriving the lower bound of the related diagnosability of a graph. Many results to this regard have appeared in literature. Readers are referred to\ \cite{CQS2017c} for the result of the $g$-good neighbor connectivity, $g \in [1, 2],$  of the arrangement graph, and its connection to its $g$-good-neighbor diagnosability, $g \in [1, 2],$ in\ \cite{CQS2019}. The $g$-good neighbor connectivity, $g \in [0, n-k],$  of the $(n, k)$-star graph is given in\ \cite[Theorem~9]{Yuan2011}, and its connection to its $g$-good-neighbor diagnosability is explored in\ \cite{CQS2019}. 
A general relationship between $g$-good-neighbor connectivity and its $g$-good-neighbor diagnosability is also discussed in\ \cite{Lin2018,Cheng2018,CQS2019}.
Moreover, the $2$-extra connectivity of the bubble-sort graph is derived in\ \cite[Theorem~3.2]{Wang2016b}, and its connection to its $g$-extra diagnosability is given in\ \cite[Theorem~5.2]{Wang2016b}. The $g$-extra connectivity of the arrangement graphs, $g=1, 2,$ and an asymptotic result for the general case, are given in\ \cite{Cheng2013}, and the $3$-extra connectivity in\ \cite{Lin2014}; and the $g$-extra diagnosability results of the arrangement graph, $g \in \{1, 2, 3\},$ are presented in\ \cite{Xu2016}. The $g$-extra connectivity result of the $(n, k)$-star graphs appeared in\ \cite{Yuan2011}, and their associated $g$-extra diagnosability is recently derived in\ \cite{Lv2019}.  It is not surprising that all these connectivity results are structure dependent, and tedious to derive. 

We will first show how to come up with an upper bound for the $M$-diagnosability of a graph $G$ via a common construction, when $M$ refers to either the $g$-good-neighbor fault-tolerant model or the $g$-extra fault-tolerant model; and then show how to derive a lower bound of its $M$ 
$t$-diagnosability once its $M$-connectivity result is available. These two bounds could lead to a  tight one when and if they  agree with each other. 

\subsection{Upper bound result derivation}
\label{section:UB}
To show that $t$ is an upper bound of $t_M(G, D),$ i.e., $t_M(G, D) \leq t,$ we only need to show that, for some pair of  distinct $M$-faulty sets, $F_1, F_2,$ $|F_1| \leq t+1, |F_2| \leq t+1,$  $V \setminus (F_1 \cup F_2) \not = \emptyset$, $(F_1, F_2)$ is  indistinguishable in $G$ according to $D.$

\begin{figure}[hb]
\begin{center}
\begin{picture}(100,60)(0,5)

\thicklines
  \put(0,0){\line(1,0){100}}
  \put(0,60){\line(1,0){100}}
  \put(0,0){\line(0,1){60}}
  \put(100,0){\line(0,1){60}}

\thinlines

  \put(50,28){\circle{20}}

  \put(20,10){\line(1,0){50}}
  \put(20,50){\line(1,0){50}}
  \put(20,10){\line(0,1){40}}
  \put(70,10){\line(0,1){40}}

  \put(50,28){\makebox(0,0){\small{$Y$}}}
  \put(30,42){\makebox(0,0){\small{$F_1$}}}
  \put(10,52){\makebox(0,0){\small{$G$}}}
  \put(90,10){\makebox(0,0){\small{$F_2$}}}
  \put(84,12){\vector(-3,2){12}}
   \end{picture}
\end{center}
\caption{A usual $M$-faulty set construction}
\label{figure:faultModel}
\end{figure}
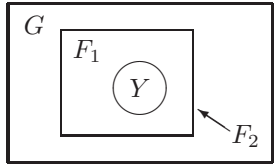

Let $G(V, E)$ be a connected graph, and $v \in V,$ we use $N(v)$ to denote the set of neighbors of $v$ in $G,$ i.e., $N_G(v)=\{w: (v, w) \in E\}.$ Let $S \subset V(G),$ we use $N_G(S)$ to denote the {\em open neighborhood of} vertices in $S,$ i.e., all the neighbors of  vertices of $S$ in $G,$ excluding those in $S$; and use $N^c_G(S)$ to denote the {\em closed neighborhood} of vertices in $S,$ that is, $N_G(S) \cup S.$ We will drop the subscript $G$ when the context is clear. A usual upper bound construction for both the $g$-good-neighbor and the $g$-extra fault-tolerant models, as shown in Figure~\ref{figure:faultModel}, is to select an non-empty set $Y\ (\subset V)$, let  $F_1=N(Y),$  and $F_2=N^c(Y),$ such that $V \not = F_2$ and both $F_1$ and $F_2$ are $M$-faulty sets. 

Since $F_1 \cup F_2=F_2,$ $F_1 \Delta F_2=Y,$   and $F_1=N(Y),$ there cannot be an edge connecting a vertex outside $F_1 \cup F_2\ (=N^c(Y)\not = V(G),$ by assumption, and any vertex in $F_1 \Delta F_2\ (=Y).$ Thus, $(F_1, F_2)$ is indistinguishable in PMC  by Theorem~\ref{theorem:PMCCase}, and MM* by Theorem~\ref{theorem:MMCase}. Finally, since $|F_2|=|N^c(Y)|,$ and $|F_1| =|N(Y)| < |N^c(Y)|,$ by the maximum restriction assumption, $t_M(G, D)  \leq |N^c(Y)|-1.$  We summarize the above discussion into the following result.
\begin{proposition}
\label{proposition:lubM}
Let $G(V, E)$ be a connected graph,  $M$ stand for either the $g$-good-neighbor or the $g$-extra fault-tolerant model, and let $Y\ (\subset V).$ If both $N(Y)$ and $N^c(Y)$ are $M$-faulty sets, and $V \setminus N^c(Y) \not = \emptyset,$ then $t_M(G, D)  \leq |N^c(Y)|-1.$
\end{proposition}
We notice that, since $F_1\ (=N(Y))$ is an $M$-faulty cut, $|N(Y)| \geq \kappa_M(G).$ In particular, we would have $t_M(G, D)  \leq t=(|Y|+\kappa_M(G))-1,$ if $N(Y)$ happens to be a minimum $M$-faulty cut. 

\begin{corollary}
\label{corollary:lubM}
Let $G(V, E)$ be a connected graph, $M$ stand for either the $g$-good-neighbor or the $g$-extra fault-tolerant model, and let $Y$ be a subset of $V.$ If both $N(Y)$ and $N^c(Y)$ are $M$-faulty sets, $N(Y)$ is a minimum $M$-faulty cut of $G,$ and $V \setminus N^c(Y) \not = \emptyset,$ then $t_M(G, D)  \leq |Y|+\kappa_M(G)-1.$ 
\end{corollary}

The following is a slightly revised version of an earlier result, by adding the necessary assumption that $V \setminus N^c(H) \not = \emptyset.$

\begin{corollary}
\label{corollary:lubG}
\cite[Theorem~3.3]{Wang2019}
Let $G(V, E)$ be a connected graph. If there is a connected sub-graph $H$ of $G$ with $|H|=g+1$ such that $N(H)$ is a minimum $g$-extra cut of $G$, and $V \setminus N^c(H) \not = \emptyset,$ then $\overline{t}_g(G, D) \leq \overline{\kappa}_g(G)+g.$ 
\end{corollary}

\noindent{\bf Proof:}  By assumption of this result, $N(H)$ is a $g$-extra faulty set. Consider a component $C$ of $V \setminus N^c(H),$ i.e., $V(C) \subset V \setminus N^c(H).$ By assumption, $V(C) \not = \emptyset.$ Clearly, $V(C) \subset V \setminus N(H).$  By the assumption that $N(H)$ is a $g$-extra faulty set, $C$ contains at least $g+1$ vertices in $V \setminus N(H).$ Since none of these vertices belong to $H,$ $N^c(H)$ is also a $g$-extra faulty set. By assumption of this result, and Corollary~\ref{corollary:lubM}, $\overline{t}_g(G, D) \leq |H|+\overline{\kappa}_g(G)-1=\overline{\kappa}_g(G)+g.$ 

\smallskip

We notice that Corollary~\ref{corollary:lubG}, although would derive a lower upper bound of the $g$-extra diagnosability of a graph, indeed its lowest upper bound in light of the forthcoming Corollaries~\ref{corollary:lbgPMC} and~\ref{corollary:ge>=2}, makes an existence assumption on a subset $H,$ which is clearly not computationally feasible to check. Hence, this result is  not effectively applicable. Theorems~1 and~2 in\ \cite{Lin2014}, on the relationship between the $g$-good-neighbor connectivity and the associated diagnosability, also share such a flavor. Such a concern leads to the following result.
\begin{corollary}
\label{corollary:lubMR}
Let $G(V, E)$ be a connected graph, and let $Y \subset V.$ If $N(Y)$ is an $M$-faulty set, where $M$ stands for either the $g$-good-neighbor or the $g$-extra fault-tolerant model, and $V \setminus N^c(Y) \not = \emptyset,$  then $t_M(G, D)  \leq |N^c(Y)|-1.$
\end{corollary}

\noindent{\bf Proof:}  In observing the proof of Corollary~\ref{corollary:lubG}, we are left to show that $N^c(Y)$ is also a $g$-good-neighbor faulty set, when $N(Y)$ is. Let $u \in V \setminus N^c(Y),$ then $u \in V \setminus N(Y).$ By definition, $u$ has at least $g$ neighbors in $V \setminus N(Y).$ Again, since $u \not \in N(Y),$ none of its neighbors could be in $Y.$ In other words, all such at least $g$ neighbors of $u$ are outside $N^c(Y),$ thus, by definition, $N^c(Y)$ is also a $g$-good-neighbor faulty set. The result now follows Proposition~\ref{proposition:lubM}. 

\smallskip
It is important to point out  that Proposition~\ref{proposition:lubM} and all its corollaries fail to apply to the conditional fault tolerant model, which requires any vertex, faulty or not, have at least one fault-free neighbor: Just consider $u \in Y,$ any of its neighbors belongs to either $Y$ or $N(Y)\ (\subset N^c(Y)).$ Since $N(Y)$ is a conditional faulty set, such a fault-free vertex $u\ (\not \in N(Y))$ must have a fault-free neighbor in $Y,$ since it could not have a neighbor in $V \setminus N^c(Y).$ On the other hand, since such a $u\ (\in N^c(Y))$ does not have a neighbor outside $N^c(Y),$ $N^c(Y)$ could not be a conditional faulty set, since this faulty vertex $u\ (\in N^c(Y))$ does not have a fault-free neighbor in terms of the faulty set $N^c(Y).$  Indeed, a different construction was made use of in deriving an upper bound for the conditional diagnosability of the hypercube structure~\cite[Lemma~11]{Lai2005}, and that for Cayley graphs generated by transposition trees\ \cite[Theorem~3]{Lin2008}. This is part of the reason that we imposed the ``{\em fault-free}'' restriction on the fault tolerant models that we study in this paper, where only a fault-free vertex has to have a property in the survival graph. In particular, the aforementioned faulty vertex $u$, as a member of the conditional faulty set $N^c(Y),$ would not need to have a neighbor outside $N^c(Y).$

\smallskip

The proof of Proposition~\ref{proposition:lubM} is rather constructive in its nature. For a construction of such a pair of $g$-good-neighbor faulty sets, $(N(Y), N^c(Y))$, in the arrangement graph, readers are referred to\ \cite[Theorem~3.1]{CQS2017c}. For a similar construction for the $(n, k)$-star graph, readers are referred to\ \cite{CQS2019,Lv2019}, for an example in the bubble-sort graph, readers are referred to\ \cite[Lemma~3.4]{Wang2016b}.
\smallskip

Recall that $t_g(G, D)$ denotes the $g$-good-neighbor diagnosability of a graph $G$ in terms of a diagnostic model $D,$ which could be either the PMC model  or the MM* model. Both the $(n, k)$-star graph\ \cite{Day1994}, denoted by $S_{n, k}, n \geq 2, k \in [1, n),$ and the arrangement graph\ \cite{Day1992}, denoted by $A_{n, k}, n \geq 2, k \in [1, n),$ are well studied interconnection networks. Their respective $g$-good-neighbor diagnosability has been obtained in\ \cite{CQS2019}; and, in this paper, we will study the $g$-extra diagnosability of these two structures in Section~\ref{section:resultNK}, and Section~\ref{section:resultA}, respectively. To start, we provide two specific upper bound results for their $g$-good-neighbor diagnosabilities, which we will make use of later in deriving the upper bound of the respective $g$-extra diagnosability of the $(n, k)$-graphs and the arrangement graphs, with the help of Corollary~\ref{corollary:gnE}.
\begin{theorem}
\cite[Theorem~5.2]{CQS2019}
\label{theorem:UBSg}
For $n \geq 4, k \in [2, n), g \in [0, n-k],$ $t_g(S_{n, k}, D) \leq n+g(k-1)-1.$ 
\end{theorem} 

\begin{theorem}
\cite[Theorem~4.3]{CQS2017c}
\label{theorem:UBAg}
For $n \geq 3, k \in [2, n), g \in [0, n-k),$ $t_g(A_{n, k}, D) \leq (n-k)[(g+1)(k-1)+1]$.
\end{theorem}

\subsection{Lower bound result derivation under the PMC model}
\label{section:LBP}

To show that $t$ is a lower bound of $t_M(G, D),$ i.e., $t_M(G, D) \geq t,$ we need to show that, for any two distinct $M$-faulty sets, $F_1, F_2,$ $|F_1| \leq t, |F_2| \leq t,$  $V \setminus (F_1 \cup F_2) \not = \emptyset$, $(F_1, F_2)$ is  distinguishable in $G$ according to $D.$ 

The following result is always used in deriving diagnosability results, as far as we know. On the other hand, it turns out that this work-horse result does not depend on either the diagnostic model, or the fault-tolerant model, as long as they are consistent with the ``fault-free'' restriction that we imposed earlier.  

\begin{proposition}
\label{proposition:intersection}
Let $G(V, E)$ be a connected graph,  $M$ stand for a fault-tolerant model, and let $F_1, F_2 \subset V.$ If both $F_1$ and $F_2$ are $M$-faulty sets in $G,$ so is $F_1 \cap F_2.$ 
\end{proposition}
\noindent{\bf Proof:} We show that if, $M,$ as a property, holds for any vertex in both $G-F_1$ and $G-F_2$\ \footnote{This is where the ``fault-free'' restriction is needed.}, then it also holds for any vertex in $G-(F_1 \cap F_2).$ 

We notice that $G-(F_1 \cap F_2)=(G-(F_1 \cup F_2)) \cup (F_1 \setminus F_2) \cup (F_2 \setminus F_1).$  Let $u \in V\setminus (F_1 \cup F_2),$ since $u$ is in $V \setminus F_1,$ $M$ holds on $u$ by assumption. Now, let $u \in F_1 \setminus F_2,$ since $M$ holds on $G-F_2,$ by definition, $M$ holds on such a $u,$ as well. The last case when $u \in F_2 \setminus F_1$ can be similarly argued. Hence, $M$ holds on any such a vertex $u,$ and $F_1 \cap F_2$ is indeed an $M$-faulty set.  

\smallskip
The following result is just one of its many specific applications. 
\begin{corollary}
\label{corollary:gCap}
Let $G(V, E)$ be a connected graph. If both $F_1$ and $F_2$ are $g$-extra faulty sets, so is $F_1 \cap F_2.$ 
\end{corollary}
\noindent{\bf Proof:} Let $C$ be a component in $F_1 \setminus F_2.$ Since it is outside of $F_2,$ a $g$-extra faulty set, it must have at least $g+1$ fault-free vertices outside $F_2,$ thus outside $F_1 \cap F_2.$ The case of $C$ being in $F_2 \setminus F_1$ can be similarly argued. Finally, Let $C$ be a component in $V \setminus (F_1 \cup F_2),$ it is outside both $F_1$ and $F_2,$ by definition, it has to contain at least $g+1$ fault-free vertices outside $F_1 \cup F_2,$ thus outside $F_1 \cap F_2,$ as well.  Hence, $F_1 \cap F_2$ is also a $g$-extra faulty set.

\smallskip

The structure independent part of the general process of deriving a lower bound of $M$-diagnosability of a graph $G$ in terms of the PMC model can be summarized with the following result.
\begin{proposition}
\label{proposition:lbPMC}
Let $G(V,E)$ be a connected graph, $\kappa_M(G)$ be the $M$-connectivity of $G$, and let $F_1, F_2$ be a pair of distinct $M$-faulty sets such that $|F_1|, |F_2| \leq t,$ and $V \setminus (F_1 \cup F_2) \not = \emptyset.$  Without loss of generality, assuming $F_1 \setminus F_2 \not = \emptyset$\ \footnote{Since $F_1 \not = F_2,$ either $F_1 \setminus F_2 \not = \emptyset$ or $F_2 \setminus F_1 \not = \emptyset.$}, if $t \geq |F_1 \setminus F_2|+\kappa_M(G)$ leads to a contradiction,  $t_M(G, \mbox{\rm PMC}) \geq t.$

\end{proposition}

\noindent{\bf Proof:}  Assume that $(F_1, F_2)$ is indistinguishable in terms of the PMC model. By Theorem~\ref{theorem:PMCCase}, let $u$ be a vertex in $V \setminus (F_1 \cup F_2),$  $u$ is not  adjacent to any vertex in $F_1 \Delta F_2.$ Since $G$ is connected, there is a path between any vertex in $V \setminus (F_1 \cup F_2)$ and another one in $F_1 \setminus F_2,$ which then has to go through a vertex in $F_1 \cap F_2.$ Hence, $F_1 \cap F_2$ must be a cut. By Proposition~\ref{proposition:intersection}, $F_1 \cap F_2$ is an $M$-faulty cut, thus  $|F_1 \cap F_2| \geq \kappa_M(G).$ Hence, 
\[
    t  \geq |F_1| = |F_1 \setminus F_2| + |F_1 \cap F_2|\geq |F_1 \setminus F_2|+\kappa_M(G).
\]
If the above inequality leads to a contradiction, by the arbitrary assumption on $F_1$ and $F_2,$ any such pair of $M$-faulty sets, $(F_1, F_2)$, must be distinguishable in $G.$ By the maximum restrictive assumption, $t_M(G, \mbox{\rm PMC}) \geq t.$ 

\smallskip

The value of the following result regarding a lower bound of the $g$-extra diagnosability of a graph $G$ is that, once its $g$-extra connectivity is available, it is effectively applicable via the condition of $|V| > 2(\overline{\kappa}_g(G)+g),$ which does depend on $G.$ 
\begin{corollary}
\label{corollary:lbgPMC}
Let $G(V, E)$ be a connected graph, and let $\overline{\kappa}_g(G)$ be its $g$-extra connectivity, $g \geq 1.$ If $|V| > 2(\overline{\kappa}_g(G)+g),$ then $\overline{t}_g(G, \mbox{\rm PMC}) \geq \overline{\kappa}_g(G)+g.$
\end{corollary}
\noindent{\bf Proof:} Let $F_1$ and $F_2$ be a pair of distinct $g$-extra faulty sets, $|F_1| \leq t=\overline{\kappa}_g(G)+g,$ and $|F_2| \leq t=\overline{\kappa}_g(G)+g.$  Assume that $(F_1, F_2)$ is indistinguishable in terms of the PMC diagnostic model. By the assumption of this result,
\begin{eqnarray*}
|V \setminus (F_1 \cup F_2)&=&|V|-|F_1 \cup F_2|=|V|-(|F_1|+|F_2|-|F_1 \cap F_2|)\\
&=&|V|-(|F_1|+|F_2|)+|F_1 \cap F_2| \geq |V|-(|F_1|+F_2|)\\
&\geq& |V|-2[\overline{\kappa}_g(G)+g] >0.
\end{eqnarray*}
Hence, $V \setminus (F_1 \cup F_2) \not = \emptyset.$

Since $G$ is connected, and, by the indistinguishable assumption, any path between a vertex in $V \setminus (F_1 \cup F_2)$ and another in $F_1 \Delta F_2$ has to go through a vertex in $F_1 \cap F_2.$ Hence, $F_1 \cap F_2$ is a cut. Since both $F_1$ and $F_2$ are assumed to be $g$-extra faulty sets, so is $F_1 \cap F_2$ by Corollary~\ref{corollary:gCap}. Thus, $F_1 \cap F_2$ is a $g$-extra cut, and $|F_1 \cap F_2| \geq \overline{\kappa}_g(G).$

On the other hand, as $F_1 \not = F_2,$ without loss of generality, assume that $F_1 \setminus F_2 \not = \emptyset.$ Since $F_1 \cap F_2$ is a cut, so is $F_2.$ Let $C$ be a component such that $V(C) \cap (F_1 \setminus F_2) \not = \emptyset.$ By the assumption that $F_1$ and $F_2$ are indistinguishable in terms of the PMC model, and Theorem~\ref{theorem:PMCCase}, no vertex in $V \setminus (F_1 \cup F_2)$ is adjacent to another vertex in $F_1 \setminus F_2$. Hence, $V(C) \subseteq F_1 \setminus F_2.$ By the assumed $g$-extra nature of $F_2,$ since $V(C) \cap F_2=\emptyset,$ $|F_1 \setminus F_2| \geq |V(C)| \geq g+1.$  


Since the assumption that $t \geq  |F_1 \setminus F_2| + \overline{\kappa}_g(G)$ would lead to the following contradiction:
\[
\overline{\kappa}_g(G)+g = t \geq |F_1 \setminus F_2| + \overline{\kappa}_g(G) \geq (g + 1) +  \overline{\kappa}_g(G)],
\]
by Proposition 3.3, ˜$\overline{t}_g(G, \mbox{\rm PMC}) \geq \overline{\kappa}_g(G)+g.$

\smallskip

Almost the same argument establishes the following result, as we notice that, under the $g$-good-neighbor circumstances, $F_1 \setminus F_2$ contains at least $g+1$ vertices since any vertex in $F_1 \setminus F_2$ has at least this many neighbors because of the $g$-good-neighbor nature of $F_2,$ and all such vertices belong to $F_1\setminus F_2$ because of the indistinguishable assumption and Theorem~\ref{theorem:PMCCase}. 
\begin{corollary}
\label{corollary:lbgnPMC}
Let $G(V, E)$ be a connected graph, and let $\kappa_g(G)$ be its $g$-good-neighbor connectivity. If $|V| > 2({\kappa}_g(G)+g),$ then ${t}_g(G, \mbox{\rm PMC}) \geq {\kappa}_g(G)+g.$
\end{corollary}

As a demonstration of the value of Proposition~\ref{proposition:lbPMC} and its corollaries, we provide an alternative derivation of the $g$-good-neighbor diagnosability of the $(n, k)$-start graph by starting with the following connectivity result.
\begin{theorem}
\label{theorem:NKconn}
\cite{Li2014}
Let $\kappa_g(G)$ be the $g$-good-neighbor connectivity of $G,$ for $n \geq 3, k \in [2, n),$ and $g \in [0, n-k],$   $\kappa_g(S_{n, k})=n+g(k-2)-1.$ 
\end{theorem}

The following result follows directly from Corollary~\ref{corollary:lbgnPMC}, and provides a concrete instance of Corollary~\ref{corollary:tg}, showing the $g$-good-neighbor diagnosability of the $(n, k)$-star graph increases monotonically in terms of $g.$ 
\begin{corollary}
\cite[Theorem~5.3]{CQS2019}
Let $t_g(G)$ be the $g$-good-neighbor diagnosability of $G,$ for $n \geq 4,k \in [2, n), g\in [0, n-k],$ $t_g(S_{n, k}, \mbox{\rm PMC}) = n+g(k-1)-1.$
\end{corollary}

\noindent{\bf Proof:} Routine arithmetic shows that $|V(S_{n, k})|=n!/(n-k)!>2[n+g(k-1)-1]=2[\kappa_g(S_{n, k})+g]$\ \cite[Theorem~5.2]{CQS2019}, when $n \geq 4,k \in [2, n).$ Notice that this condition leads to a restriction on the parameters $n$ and $k.$ Thus, by Corollary~\ref{corollary:lbgnPMC}, $
t_g(S_{n, k}, \mbox{\rm PMC}) \geq n+g(k-1)-1,$ which is actually a tight bound, thanks to Theorem~\ref{theorem:UBSg}.  
\bigskip

We remark that the above process is much shorter, and cleaner, than the original one, as shown in\ \cite{CQS2019}.

\subsection{Lower bound result derivation under the MM* model}
\label{section:LBM}
The process of deriving a lower bound result of the $M$-diagnosability in terms of the MM* diagnostic model is essentially the same as that for the PMC model, except that we also need to show that no isolated vertex exists in the non-empty $V(G) \setminus (F_1 \cup F_2),$ where $(F_1, F_2)$ is the pair of indistinguishable $M$-faulty sets that we use to construct the desired contradiction as required in Proposition~\ref{proposition:lbPMC}.  The reason for this additional requirement, the ``{\em isolation condition}'' henceforth, is that, in this MM* case, if $u$ is isolated in $V(G) \setminus (F_1 \cup F_2),$ it can be adjacent to some vertex in $F_1 \Delta F_2.$ Then, $F_1 \cap F_2$ would not be a cut, thus the reasoning as we followed earlier in establishing Proposition~\ref{proposition:lbPMC} is no longer applicable. 
\begin{proposition}
\label{proposition:lbMM}
Let $G(V,E)$ be a connected graph, $\kappa_M(G)$ be the $M$-connectivity of $G$, and let $F_1, F_2$ be two $M$-faulty sets, such that $|F_1|, |F_2| \leq t,$ and the non-empty set of $V \setminus (F_1 \cup F_2)$ contains no isolated vertex. Assume that $F_1 \setminus F_2 \not = \emptyset$\ \footnotemark, if $t \geq |F_1 \setminus F_2|+\kappa_M(G)$ leads to a contradiction, then $t_M(G, \mbox{\rm MM*}) \geq t.$
\end{proposition}
\footnotetext{For a given pair of supposedly indistinguishable  pair of distinct $M$-faulty set, $F_1, F_2,$ If $F_1 \setminus F_2 =\emptyset,$ then $F_1 \subset F_2.$ 
   \begin{itemize}  
     \item If both $F_1$ and $F_2$ are $g$-good-neighbor faulty sets, then when $F_1 \subset F_2,$ any vertex, $u,$  in $V(G) \setminus (F_1 \cup F_2)$ must have at least $g$ neighboring vertices outside $F_2,$ thus such a vertex $u$ has at least $g\ (\geq 1)$ neighbors in $V(G) \setminus (F_1 \cup F_2),$ i.e., it is not isolated. This would lead to a simpler argument as given in Proposition~\ref{proposition:lbPMC}.
     \item If both $F_1$ and $F_2$ are $g$-extra faulty sets, then when $F_1 \subset F_2,$ any component in $V(G) \setminus (F_1 \cup F_2)$ much have at least $g+1$ vertices $F_2,$ thus any vertex $u$ in $V(G) \setminus (F_1 \cup _2)$ has at least one neighbor in $V(G) \setminus (F_1 \cup F_2),$ i.e., it is not isolated, either.
   \end{itemize}      
Thus, we can assume that $F_1 \setminus F_2 \not =\emptyset.$ Moreover, since $F_1 \not = F_2,$ if $F_1 \setminus F_2 \not =\emptyset,$ then $F_2 \setminus F_1 \not =\emptyset.$ }
\noindent{\bf Proof:} Assume that $(F_1, F_2)$ is indistinguishable in $G$ in terms of the MM* diagnostic model. Let $u \in V \setminus (F_1 \cup F_2),$ by the assumed ``isolation condition'', $u$ is not isolated, thus adjacent to another vertex $w \in V \setminus (F_1 \cup F_2).$ By the indistinguishable assumption and Theorem~\ref{theorem:MMCase}, $u$ is not adjacent to any vertex in $F_1 \Delta F_2.$ Thus, $F_1 \cap F_2$ is a cut by the assumption that $G$ is connected.  Together with Proposition~\ref{proposition:intersection}, $F_1 \cap F_2$ is a $M$-faulty cut. Hence, $|F_1 \cap F_2| \geq \kappa_M(G).$

By assumption of this result,
\[
    t  \geq |F_1| = |F_1 \setminus F_2| + |F_1 \cap F_2|=|F_1 \setminus F_2|+\kappa_M(G).
\]
If the above inequality leads to a contradiction, then, any such pair of $M$-faulty sets, $(F_1, F_2)$, must be distinguishable in $G.$ By the assumed maximum restriction, $t_M(G, \mbox{\rm MM*}) \geq t.$ 

\smallskip

It turns out that this additional isolation condition is not needed when $g \geq 2$ for both $g$-good-neighbor and $g$-extra fault tolerant models as shown in the following Corollaries~\ref{corollary:g>=2} and~\ref{corollary:ge>=2}, respectively.
\begin{corollary}
\label{corollary:g>=2}
Let $G(V,E)$ be a connected graph, and let $\kappa_g(G), g \geq 2,$ be the $g$-good-neighbor connectivity of $G.$ If $|V|>2(\kappa_g(G)+g),$ then $t_g(G, \mbox{\rm MM*}) \geq \kappa_g(G)+g.$ 
\end{corollary}
\noindent{\bf Proof:} Let $F_1, F_2$ be two $g$-good-neighbor faulty sets, $|F_1| \leq t=\kappa_g(G)+g,$ $|F_2| \leq t=\kappa_g(G)+g.$ Assume that $(F_1, F_2)$ is indistinguishable. 

The same argument as made in proving Corollary~\ref{corollary:lbgPMC} shows that the condition of $|V|>2(\kappa_g(G)+g)$ implies that $V \setminus (F_1 \cup F_2) \not = \emptyset.$ 

Let $w$ be a vertex in $V \setminus (F_1 \cup F_2).$  Since $F_1$\ (respectively, $F_2$) is a $g$-good-neighbor faulty set, $w$ has at least $g\ (\geq 2) $ neighbors outside $F_1$\ (respectively, $F_2$). Since  $F_1$ and $F_2$ are indistinguishable, $w$  will have at most one neighbor in $F_2 \setminus F_1$\ (respectively, $F_1 \setminus F_2$). Thus, it has at least $g-1\ (\geq 1)$ neighbor(s) in $V \setminus (F_1 \cup F_2).$ In other words, $w$ could not be isolated in $\overline{F_1 \cup F_2}$. 

The same argument as made in proving Corollary~\ref{corollary:lbgnPMC} shows that $|F_1 \setminus F_2| \geq g+1.$ The result now follows from Proposition~\ref{proposition:lbMM} since the assumption that 
\[
\kappa_g(G)+g =t \geq |F_1 \setminus F_2|+\kappa_g(G)\geq g+1+\kappa_g(G)
\]
would lead to  a contradiction, showing that such a pair of $g$-good-neighbor faulty sets must be distinguishable, and the result follows from the ``maximum restriction''.

\begin{corollary}
\label{corollary:ge>=2}
Let $G(V,E)$ be a connected graph, and let $\kappa_g(G), g \geq 2,$ be the $g$-extra connectivity of $G.$ If $|V|>2(\overline{\kappa}_g(G)+g),$ then $\overline{t}_g(G, \mbox{\rm MM*}) \geq \overline{\kappa}_g(G)+g.$ 
\end{corollary}
\noindent{\bf Proof:}  Let $F_1, F_2$ be two $g$-extra faulty sets, $|F_1| \leq t=\overline{\kappa}_g(G)+g,$ $|F_2| \leq t=\overline{\kappa}_g(G)+g.$ Assume that $(F_1, F_2)$ is indistinguishable. Let $w \in V \setminus (F_1 \cup F_2)$ and let $C$ the component that contains $w.$  By the indistinguishable nature of $F_1$ and $F_2,$ $|V(C) \cap (F_1 \setminus F_2)| \leq 1,$ and $|V(C) \cap (F_2 \setminus F_1)| \leq 1.$ Furthermore, by the $g$-extra assumption on both $F_1$ and $F_2,$ $C$ contains at least $g+1 \geq 3$ vertices outside $F_1$ and $F_2.$ Hence, $C$ would have to contain at least another vertex in $V \setminus (F_1 \cup F_2).$ Thus, $w$ is not isolated, either.   

The same argument as made in proving Corollary~\ref{corollary:lbgPMC} shows that $|F_1 \setminus F_2| \geq g+1.$ The rest of the proof is the same as that for the above result. 

\smallbreak

We give an example in this regard as follows:
\begin{theorem}
\label{theorem:Ag=2}
\cite{CQS2017c}
For $n \geq 8, \kappa_2(A_{n, 2}) =4n-12;$ and, for $k \in [3, n-5]\cup \{n-2,n-1\},$ $\kappa_2(A_{n, k}) = (3k-2)(n-k)-2.$ 
\end{theorem}

\begin{corollary}
\label{corollary:glbgAMM}
\cite{CQS2019}
For $n \geq 7, k \in [4, n-1), t_2(A_{n, k}, \mbox{\rm MM}^*) = (3k-2)(n-k).$
\end{corollary}

\noindent{\bf Proof:} Routine arithmetic shows that $|V(A_{n, k})|=n!/(n-k)!>2[(3k-2)(n-k)]$\ \cite[Theorem~4.4]{CQS2019}, when $n \geq 7, k \in [4, n-1].$  By Corollary~\ref{corollary:g>=2} and Lemma~\ref{theorem:Ag=2}, $t_2(A_{n, k}, \mbox{\rm MM}^*) \geq (3k-2)(n-k),$ which is actually a tight bound by Theorem~\ref{theorem:UBAg}, taking $g=2.$  

\smallskip

We give another example, where we have to enforce the isolation condition  when $g=1.$ We start with the important connectivity result. 
\begin{theorem}
\label{theorem:g=1}
\cite{CQS2017c}
 For $n \geq 3, n \not = 4, k \in [2, n),$ $\kappa_1(A_{n, k}) = (2k-1)(n-k)-1.$ And $\kappa_1(A_{4, 2}) = \kappa_1(A_{4, 3})\ (=\kappa_1(S_{4}))= 4.$
\end{theorem}

The following provides the needed no-isolation-vertex result. 
\begin{lemma}
\label{lemma:A}
\cite{CQS2019}
Let $F_1, F_2$ be two distinct $1-$good-neighbor conditional cuts of $A_{n, k}, n \geq 6, k \in [5, n-1)$, or $n \geq 11, k \in [10, n),$ $|F_1|, |F_2| \leq (2k-1)(n-k),$  such that $V \setminus (F_1 \cup F_2) \not = \emptyset.$ Then, $V \setminus (F_1 \cup F_2)$ contains no isolated vertices.
\end{lemma}

We are now ready to achieve the following lower bound result. 
\begin{corollary}
\cite{CQS2019}
\label{theorem:mmlbg=1}
For $n \geq 6, k \in [5, n-1)$, or $n \geq 11, k \in [10, n),$ $t_1(A_{n, k}, {\rm MM}^*)=(2k-1)(n-k).$
\end{corollary}

\noindent{\bf Proof:} Let $F_1, F_2$ be two distinct $1$-good-neighbor faulty sets, $|F_1|, |F_2| \leq t=(2k-1)(n-k),$ and assume that $(F_1, F_2)$ is indistinguishable in terms of the MM* model. 

Routine arithmetic shows that $|V(A_{n, k})|=n!/(n-k)!>2[(2k-1)(n-k)]=2[\kappa_2(A_{n, k})+1]$, when $n \geq 5, k \in [2, n).$ Thus, $|V(A_{n, k})/(F_1 \cup F_2)|$ does not contain an isolated vertex by Lemma~\ref{lemma:A}.  

Without loss of generality, $F_1 \setminus F_2 \not = \emptyset.$ By an argument similar to that made in Corollary~\ref{corollary:lbgPMC}, $|F_1 \setminus F_2| \geq 2.$ Since $(2k-1)(n-k)=t \geq |F_1 \setminus F_2|+\kappa_1(G) \geq (2k-1)(n-k)-1+2=(2k-1)(n-k)+1$ is a contradiction, by Proposition~\ref{proposition:lbMM}, $t_1(A_{n, k}, {\rm MM}^*) \geq (2k-1)(n-k),$  which is again a tight bound by Theorem~\ref{theorem:UBAg}, taking $g=1.$  
\bigskip

To recapitulate,  a uniform construction of an appropriate faulty cut in a graph $G$ leads to an upper bound of both the $g$-good-neighbor, and $g$-extra, diagnosability of $G,$ in terms of both the PMC model and the MM* model, as shown in Corollary~\ref{corollary:lubMR}.  On  the other hand, the size of a minimum $g$-extra faulty cut of  $G,$ referred to as its {\em $g$-extra connectivity,} plays a critical role to derive a lower bound of its $g$-extra diagnosability, as shown in  Corollary~\ref{corollary:lbgPMC} for the PMC model, and in Corollary~\ref{corollary:ge>=2} for the MM* model. The same situation arises for the $g$-good-neighbor diagnosability, as shown in Corollary~\ref{corollary:lbgnPMC} for the PMC case, and in Corollary~\ref{corollary:g>=2} for the MM* case. 

When, and if, these two bounds agree, we will obtain the exact bound of the desired $M$ diagnosability of the graph $G$ in terms of a certain diagnostic model $D.$ In the rest of this paper, we will apply this general process to derive the $g$-extra diagnosabilities of the hypercube, the $(n, k)$-star, and the arrangement graph.

\section{The $g$-extra diagnosability of the hypercube graph}
\label{section:resultQ}

The hypercube, $Q_n$, \ \cite{Harary1988} is perhaps one of the most studied, also the simplest, interconnection networks, with commercial applications\ \cite{Hillis1985,Hayes1986}. It is $n$-regular, both vertex and edge transitive, with small diameter. Several hypercube variants have also been suggested, including augmented cubes, crossed cubes, enhanced cubes, folded cubes, m\"{o}bius cubes, twisted cubes, and (generalized) exchanged cubes. Many algorithms have been designed to run on these hypercube based architectures to solve realistic issues in applications. As a recent example, Bcube, a general hypercube based structure, was suggested in\ \cite{Guo2009} as a network structure to support reconfigurable modular data centers.

The $g$-extra diagnosability of the hypercube structure, in terms of  both the PMC and the MM* models, have been derived earlier\ \cite{Zhang2016,Zhu2016,Liu2017}   by following a structure dependent derivation process. As an opening example, we will show how to follow the general process that we discussed in the previous section to derive the $g$-extra diagnosability of the hypercube structure, denoted by $Q_n,$ for $ n \geq 4,$ and $g \in [1, n-3].$ We will also explore its $g$-extra diagnosability for a wider range of $n,$ making use of some recent results on its $g$-extra connectivity.

Let $K_2$ be the complete graph with two vertices 0, and 1; and let `$\Box$' be the Cartesian product,  $Q_n$,  $n\  (\geq 2),$  can be defined as follows:
\begin{eqnarray*}
Q_1 &=& K_2,\\
\mbox{\rm for all $n \geq 2,$ } Q_n &=& K_2 \Box Q_{n-1} .
\end{eqnarray*}

Thus, a vertex of $Q_n,$ $u,$ can be represented as an $n$-bit binary string: $(u_0, u_1, \ldots, u_{n-1}),$ where, for all $i \in [0, n-1], u_i \in \{0, 1\}.$ Clearly, $Q_n$ contains $2^n$ vertices, and two vertices of $Q_n$ are adjacent to each other if and only if their corresponding binary strings differ in exactly one position. 

We first seek an upper bound of $\overline{t}_g(Q_n, D),$ the $g$-extra diagnosability of $Q_n, n \geq 4,$ in terms of a diagnostic model, $D,$ which could be either the PMC model or the MM* model, through the usual construction, originally suggested in\ \cite[Theorem~4.3]{CHQS2019}. 

Let $Y$ be a star graph $K_{1, g},$ $g \in [0, n-3], n \geq 4,$ consisting of $g+1$ vertices $u_0, u_1, \ldots, u_g$ such that $u_0=(0, \ldots, 0), $ $u_1=(0, 1, \ldots, 0),$   $\ldots,$ and $u_g=(\overbrace{0, 0, \ldots, 0}^g, 1, 0, \ldots, 0).$ By definition, for all $ i \in [1, g], $ $u_0$ is adjacent to $u_i;$ for all $ i, j  \in [1, g], $ $u_i$ is not adjacent to $u_j;$ and, for all $i, j \in [1, g], i < j, $ $u_i$ and $u_j,$  share exactly two common neighbors: $u_0$ and $(\underbrace{\overbrace{0, \ldots, 0}^i, 1, 0, \ldots, 0}_j, 1, 0, \ldots, 0).$ 

Since $u_0$ has $n-g$ neighbors other than those in $Y;$  each of the $g$ vertices $u_i, i \in [1, g],$ has $n-1$ neighbors in $Q_n-Y,$ but  each of  ${g \choose 2}$ pairs of them shares a common neighbor, we have the following
\begin{eqnarray}
|N(Y)|&=& g(n-1)-{g \choose 2}+(n-g) \nonumber  \\
&=&\frac{1}{2}(g+1)[2(n-1)-g]+1, \mbox{ \rm and,}\label{eq:ny} \\
|N^c(Y)|&=&|N(Y)|+|Y|=\frac{1}{2}(g+1)[2(n-1)-g]+1+(g+1) \nonumber \\
&=& \frac{1}{2}(g+1)(2n-g)+1. \label{eq:nc}
\end{eqnarray}

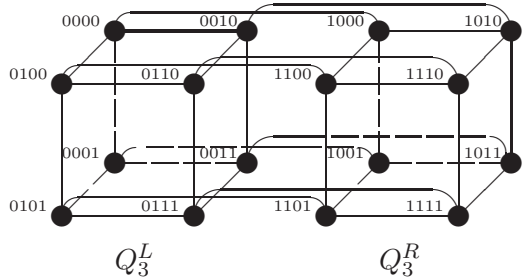
\begin{figure}[ht]
\begin{center}
\begin{picture}(140,90)(0,0)
  \put(0,0){\circle*{8}}
  \put(50,0){\circle*{8}}
  \put(0,50){\circle*{8}}
  \put(50,50){\circle*{8}}
  \put(100,0){\circle*{8}}
  \put(150,0){\circle*{8}}
  \put(100,50){\circle*{8}}
  \put(150,50){\circle*{8}}
  
  \put(20,20){\circle*{8}}
  \put(70,20){\circle*{8}}
  \put(20,70){\circle*{8}}
  \put(70,70){\circle*{8}}
  \put(120,20){\circle*{8}}
  \put(170,20){\circle*{8}}
  \put(120,70){\circle*{8}}
  \put(170,70){\circle*{8}}

  
  \thinlines
 
   \put(0,0){\line(1,0){50}}
    \put(0,0){\line(0,1){50}}
    \put(0,50){\line(1,0){50}}
    \put(50,0){\line(0,1){50}}
   \put(0,50){\line(1,1){20}}
   \put(50,50){\line(1,1){20}}
   \put(20,70){\line(1,0){50}}
   \put(50,00){\line(1,1){20}}
    \put(70,20){\line(0,1){50}}
    \put(0,0){\line(1,1){10}}
    \put(12,12){\line(1,1){10}}
    
    \put(20,22){\line(0,1){10}}
    \put(20,34){\line(0,1){10}}
    \put(20,46){\line(0,1){10}}
     \put(20,58){\line(0,1){10}}
     
     \put(22,20){\line(1, 0){10}}
     \put(34,20){\line(1, 0){10}}
     \put(46, 20){\line(1, 0){10}}
     \put(58, 20){\line(1, 0){10}}
  
   
    \put(100,0){\line(1,0){50}}
    \put(100,0){\line(0,1){50}}
    \put(100,50){\line(1,0){50}}
    \put(150,0){\line(0,1){50}}
   
   \put(100,50){\line(1,1){20}}
   \put(150,50){\line(1,1){20}}
   \put(120,70){\line(1,0){50}}
    \put(150,00){\line(1,1){20}}
    \put(170,20){\line(0,1){50}}
     \put(100,0){\line(1,1){10}}
    \put(112,12){\line(1,1){10}}
    
     \put(120,22){\line(0,1){10}}
    \put(120,34){\line(0,1){10}}
    \put(120,46){\line(0,1){10}}
     \put(120,58){\line(0,1){10}}
     
     \put(122,20){\line(1, 0){10}}
     \put(134,20){\line(1, 0){10}}
     \put(146, 20){\line(1, 0){10}}
      \put(158, 20){\line(1, 0){10}}

   \put(70,70){\oval(100,14)[t]}
   \put(120,70){\oval(100,20)[t]}
   
   \put(50,50){\oval(100,14)[t]}
   \put(100,50){\oval(100,20)[t]}
   
   
  \put(50,0){\oval(100,14)[t]}
   \put(100,0){\oval(100,20)[t]}
   
    \put(28,20){\oval(12,12)[tl]}    
   \put(32, 26){\line(1, 0){10}}
   \put(44, 26){\line(1, 0){10}}
    \put(56, 26){\line(1, 0){10}}
    \put(68, 26){\line(1, 0){30}}
    \put(100, 26){\line(1, 0){10}}
     \put(112,20){\oval(12,12)[tr]}

   \put(100,20){\oval(60,20)[tl]}    
    \put(102, 30){\line(1, 0){10}}
    \put(114, 30){\line(1, 0){10}}
     \put(126, 30){\line(1, 0){10}}
     \put(138, 30){\line(1, 0){10}}
      \put(150, 30){\line(1, 0){10}}
     \put(162,20){\oval(12,20)[tr]}    

\put(0, 72){\tiny 0000} 
\put(-20, 52){\tiny 0100} 
\put(52, 72){\tiny 0010} 
\put(30, 52){\tiny 0110} 
\put(-20, 2){\tiny 0101} 
\put(30, 2){\tiny 0111} 
\put(52, 22){\tiny 0011} 
\put(0, 22){\tiny 0001} 
\put(100, 72){\tiny 1000} 
\put(80, 52){\tiny 1100} 
\put(152, 72){\tiny 1010} 
\put(130, 52){\tiny 1110} 
\put(80, 2){\tiny 1101} 
\put(130, 2){\tiny 1111} 
\put(152, 22){\tiny 1011} 
\put(100, 22){\tiny 1001} 

\put(20, -20){$Q_3^L$}
\put(120, -20){$Q_3^R$}


\thicklines

  \put(20, 70){\line(1, 0){50}}

 \end{picture}
\end{center}

\caption{A $K_{1, 2}$ in $Q_4$} 
\label{figure:q4}
\end{figure}

For example, in a $Q_4$, as shown in Figure~\ref{figure:q4},   if we choose $K_{1, 2},$ consisting of $u_0=0000$ and $u_1=0010,$ then $Y=\{u_0, u_1\}.$ Since $u_0$ is also adjacent to $0001, 0100,$ and $1000,$  $u_1$  is adjacent to three other vertices: 0110, 0011, and 1010, $N(Y)$ contains these six vertices, while $N^c(Y)$ contains two more vertices in $Y.$ Indeed, Eqs.~\ref{eq:ny} and~\ref{eq:nc} return 6 and 8, respectively.

W need to show that $N(Y)$ is a $g$-extra faulty set, i.e., every component in $Q_n-N(Y)$ contains at least $g+1$ vertices, and $|V(Q_n)|> |N^c(Y)|$, so that we can apply Corollary~\ref{corollary:lubMR} to obtain 
\begin{eqnarray}
\overline{t}_g(Q_n, D) \leq |N^c(Y)|-1=\frac{1}{2}(g+1)(2n-g). \label{eq:QUB}
\end{eqnarray}
  
To this regard, we decompose $Q_n$ to $Q_{n-1}^L$ and $Q_{n-1}^R,$ such that $Q_{n-1}^L$\ (respectively,  $Q_{n-1}^R$) contains vertices $(0, a_1, \ldots, a_{n-1})$\  (respectively, $(1, a_1, \ldots, a_{n-1})$) where, for all $i \in [1, n-1], a_i \in \{0, 1\}.$ It is clear that, both $Q_{n-1}^L$ and $Q_{n-1}^R$ are isomorphic to $Q_{n-1};$   and each vertex in  $Q_{n-1}^L$ is associated with a unique vertex in  $Q_{n-1}^R.$ In particular, each of the $g+1$ vertices, $u_i, i \in [0, g],$ in  $Y\ (\subset V(Q_{n-1}^L))$ has a unique neighbor $u'$ in $Q_{n-1}^R.$  For example,  as shown in Figure~\ref{figure:q4}, $u_0'=1000,$  and $u_1'=1010.$

As the connectivity of $Q_{n-1}^R$ is $n-1,$ and $n-1-(g+1)=(n-2)-g \geq  (n-2)-(n-3) \geq 1,$ $Q_{n-1}^R-N^c(Y)$ is connected\ \footnote{We comment that it is this line of reasoning that requires $g \leq n-3.$  On the other hand, this condition is only sufficient. For example, if we choose a $K_{1, 2}$ in $Q_4$ as shown in Figure~\ref{figure:q4}, when $g=2>1=n-3,$  with $u_0=0000, u_1=0010,$ and $u_2=0100,$ we would still have the result that $Q_{n-1}^R -N(Y)$ is connected, thus, $Q_n-N^c(Y)$ contains exactly two components.  The upper bound of $g$ can be further expanded via alternative constructions\ \cite[Section~3]{Zhu2016b}.}, and each vertex in $Q_{n-1}^L-N^c(Y)$ is connected to its neighbor in $Q_{n-1}^R-N^c(Y).$ For example, in Figure~\ref{figure:q4}, $Q_{n-1}^L-N^c(Y)$ consists of two vertices, 0101 and 0111, both of which are adjacent to their respective neighbor in  $Q_{n-1}^R-N^c(Y).$ Thus, $Q_n-N(Y)$ consists of exactly two components, $Y$ and $Q_n-N^c(Y).$  
  
  Since $|Y|=g+1,$ we only need to show that $|Q_n-N^c(Y)|\geq g+1,$ i.e., by Eq.~\ref{eq:nc},
\begin{eqnarray*}
2^n- \left [\frac{1}{2}(g+1)(2n-g)+1 \right ]\geq g+1, \mbox{ \rm i.e.,}
\end{eqnarray*}
 \[
2^n \geq \frac{1}{2}(g+1)(2n-g)+1+(g+1)=\frac{1}{2}(g+1)(2n+2-g)+1. \]
Since $g \in [1, n-3],$ the above holds if the following does:
 \[
 2^n \geq \frac{1}{2}(n-2)(2n+1)+1,
 \]
 which holds for all $n \geq 2.$ Thus, Eq.~\ref{eq:QUB} holds, since, as a byproduct, $|V(Q_n)-N^c(Y)|\geq g+1\geq 2,$ thus, $|V(Q_n)|> |N^c(Y)|,$ as required by Corollary~\ref{corollary:lubMR}.
  
 We now turn to the lower bound of $\overline{t}_g(Q_n, D) $. As mentioned earlier, the $M$-connectivity of a graph is the key to achieving the lower bound of its $M$-diagnosability, which is often challenging and tedious. We now describe an alternative approach for its derivation through the super-connectedness of the involved graph. A graph is {\em super $m$-vertex connected of order $q$} if, with at most $m$ vertices being deleted, the survival graph is either connected or it consists of a large component and all the small components contain at most $q$ vertices\ \cite{Yuan2011,Cheng2013,Zhu2016,CQS2017}. This super-connectedness based property has also recently been applied to derive $g$-extra connectivity of the arrangement graph\ \cite{Xu2016} and the $(n, k)$-star graph\ \cite{Lv2019}. For a detailed discussion about this revealing structural property of a graph, and its close relationship to various fault-tolerant properties, including $g$-good-neighbor connectivity, $g$-extra connectivity, component connectivity, cyclic connectivity, as well as Menger connectedness, readers are referred to\ \cite{CHQS2019}.   In particular,  a set of vertices $F$ in a connected non-complete graph  $G$ is called a {\em restricted vertex-cut of order $m$} if  the survival graph $G - F$ is disconnected and every component in $G -F$ has at least $m$ vertices\ \cite{CHQS2019}. It is clear that a $g$-extra faulty set of such a graph $G$ is simply a restricted vertex-cut of order $g+1.$ The following result naturally follows the definition.
 
\begin{theorem}
 \cite{CHQS2019}
Let $G$ be an $r$-regular graph. If $G$ is super $p$-vertex-connected of order $q,$ then the restricted vertex connectivity of order $q + 1$ is at least $p + 1.$
 \end{theorem}
Since, for $n \geq 4$ and $k \in [1, n-2],$  $Q_n$ is super $(kn- k(k + 1)/2)$-vertex connected of order $k-1$\  \cite{Yang2006}, we immediately have the following result, which first appeared in\ \cite{Yang2009}.
 \begin{corollary}
 \cite{CHQS2019}
Let $n\geq 4$ and $k \in [1, n-2],$  the restricted vertex connectivity of order $k$ of $Q_n$ is $[kn-k(k+1)/2]+1.$
\end{corollary}
 
 Setting $k=g+1, $ we  have the following result, after simplification. 
 \begin{corollary}
 \label{corollary:kQ}
 Let $n\geq 4$ and $g \in [0, n-3],$  $
 \overline{\kappa}_g(Q_n)=\frac{1}{2}(g+1)[2(n-1)-g]+1.$
 \end{corollary}
 We are now ready to derive the $g$-extra diagnosability of $Q_n$ with the PMC model.
  \begin{theorem}
 \label{theorem:tQPMC}
 Let $n\geq 4$ and $g \in [1, n-3],$  $
\overline{t}_g(Q_n, \mbox{\rm PMC}) =\frac{1}{2}(g+1)(2n-g). 
 $
\end{theorem}
 \noindent{\bf Proof:} By Eq.~\ref{eq:QUB}, and Corollary~\ref{corollary:lbgPMC}, we only need to prove $2^n>(g+1)(2n-g),$ which holds if 
 \[
 2^n > (n-2)(2n-1).
 \]
 It is certainly true when $n \geq 2.$  
 \smallskip
 
The above result slightly generalizes the one achieved in\ \cite[Theorem~3.11]{Zhang2016}, when $n\geq 4, g \in [0, n-4].$  

Given a graph $G(V, E),$ and let $Y \subset V,$ the {\em vertex boundary number of} $Y$ is simply $|N(Y)|,$  denoted by $b_v(H; G)$; and the {\em minimum $k$-boundary number of} $G$ is defined as the {\em minimum boundary number of all its subgraphs with order $k,$ denoted
by $b_v(k; G).$} The relationship between $\overline{\kappa}_g(Q_n)$, the $g$-extra connectivity of $Q_n,$  and $b_v(k, Q_n)$ is recently explored in\ \cite{Zhu2016b}. As a result,  $\overline{\kappa}_g(Q_n)$ is derived for  a much bigger range of $g \in [0, 3n-7]$.
\begin{theorem}
\label{theorem:kQExtended}
\ \cite{Zhu2016b}. 
\begin{eqnarray*}
\overline{\kappa}_g(Q_n)=\left \{
\begin{array}{ll}
-\frac{1}{2} {(g+1)}^2+(n-\frac{1}{2})(g+1)+1 & n \geq 5, g \in [0, n-4]  \\
-\frac{1}{2}{(n-2)}^2+(n-\frac{1}{2})(n-2)+1 & n \geq 5, g \in [n-3, n]  \\
-\frac{1}{2} {(g+1)}^2+(2n-\frac{3}{2})(g+1)-n^2+2 & n \geq 7, g \in [n+1, 2n-5]  \\
-\frac{1}{2} {(2n-3)}^2+(2n-\frac{3}{2})(2n-3)-n^2+2 & n \geq 7, g \in [2n-4, 2n-1],\ \mbox{\rm and,}  \\
-\frac{1}{2} {(g+1)}^2+(3n-\frac{7}{2})(g+1)-3n^2+4n+2& n \geq 9, g \in [2n, 3n-7].
\end{array}
\right.
\end{eqnarray*}
\end{theorem}

We notice that, for $n \geq 5, g \in [0, n], $ this extended $\overline{\kappa}_g(Q_n)$ agrees with the result as shown in Corollary~\ref{corollary:kQ}. 
 This extended $g$-extra connectivity result has also been used to derive the following  $g$-extra diagnosability result for $Q_n$.
\begin{theorem}
\label{theorem:atQ}
\cite{Zhu2016}
Let  $n \geq 9, $   $g \in [0, 3n-7],$   then $\overline{t}_g(Q_n, \mbox{\rm PMC})=\overline{\kappa}_g(Q_n)+g.$
\end{theorem}
Similarly, by Eq.~\ref{eq:QUB}, and Corollary~\ref{corollary:ge>=2}, we have the following $g$-extra diagnosability result for $Q_n$ in terms of the MM* model.
  \begin{corollary}
 Let $n \geq 4$ and $g \in [2, n-3],$  $\overline{t}_g(Q_n, \mbox{\rm MM*}) =\frac{1}{2}(g+1)(2n-g). $
 \end{corollary}
 Regarding the missing case of $g=1$ in the above result, we notice that it is shown in\ \cite[Theorem~3.6]{Liu2017} that, for $n \geq 5,$ and $g \in [0, n-3],$
 \[
 t_g(Q_n, \mbox{ \rm MM*})=(n-g+1)2^g+1.
 \]
 By Corollary~\ref{corollary:g1=tg1}, we have that for $n \geq 5,$
 \[
 \overline{t}_1(Q_n, \mbox{\rm MM*})=t_1(Q_n, \mbox{ \rm MM*})=2n-1.
 \]
  
It has been shown, in\ \cite{Liu2017}, that $\overline{t}_1(Q_3, \mbox{ \rm MM*})=3.$ It is also stated in\ \cite[Proposition~3.8]{Liu2017} that $\overline{t}_1(Q_4, \mbox{ \rm MM*})=5,$ We present a different strategy here to demonstrate our general approach.

In place of the example as given in\ \cite[Figure~5]{Liu2017}, showing an indistinguishable pair of 1-extra faulty sets $(F_1, F_2),$  a correct one should be $F_1=\{0000, 0101, 0011, 1100, 1010, 1111\},$ and $F_2=\{0110, 0101, 0011, 1100, 1010,$ $ 1001\}.$ Indeed, as shown in Figure~\ref{figure:q4}, since there do not exist adjacent vertices $u$ and $v$ in $Q_4-(F_1 \cup F_2)$ and no vertex in $Q_4-(F_1 \cup F_2)$ is adjacent to two vertices in either $F_1 \setminus F_2$ or $F_2 \setminus F_1,$ this example does show that $\overline{t}_1(Q_4) \leq 5.$

To prove that $\overline{t}_g(Q_4) \geq 5,$ we could make use of the following super-connectedness property of $Q_n.$
\begin{theorem}
\label{theorem:sQ4}
\cite{Yang2006} If $n \geq 4$ with $k \in [1, n-1],$ then $Q_n$ is super $kn-k(k + 1)/2$-vertex connected of order $k-1.$
\end{theorem}
In other words, let $F \subset V(Q_n),$ $|F| \leq kn-k(k + 1)/2,$ then    either $Q_n-F$ is connected; or $Q_n-F$ contains a large component and all the smaller components contains at most $k-1$ vertices. Equivalently, if all such smaller components contain, in particular, any of them contains, at least $k$ vertices, then $|F| \geq [kn-k(k + 1)/2]+1.$

Now let $F_1,  F_2$ be any distinct 1-extra faulty sets such that $|F_1|, |F_2| \leq 5.$  By assumption, since $|F_1 \cup F_2| \leq 10,$ $V(Q_4) \setminus (F_1 \cup F_2) \not = \emptyset.$  Let $w$ be any vertex in $Q_4-(F_1 \cup F_2),$ it is shown in\ \cite[Proposition~3.8]{Liu2017} that $w$ is not isolated in $Q_4-(F_1 \cup F_2)$. If $(F_1, F_2)$ is not distinguishable, By Theorem~\ref{theorem:MMCase},  $w$ is not adjacent to any vertex in $F_1 \Delta F_2. $   Since $Q_4$ is connected, $F_1 \cap F_2$ must be a cut. 

Furthermore, by  assumption, both $F_1$ and $F_2$ are 1-extra faulty sets,  so is $F_1 \cap F_2$ by Corollary~\ref{corollary:gCap}. Let $C, |C| \geq 2, $ be a minimum component of $Q_4-(F_1 \cap F_2).$  Since $Q_4-(F_1 \cap F_2)$ is disconnected, by Lemma~\ref{theorem:sQ4}, taking $n=4, k=2,$  to have such a component $C$ containing at least two vertices, $ |F_1 \cap F_2| \geq 6.$  On the other hand, since $F_1 \not =  F_2,$ without loss of generality, assume that $|F_1 \setminus F_2| \geq 1,$ then $6  \leq |F_1 \cap F_2|=|F_1|-|F_1\setminus F_2| \leq 4,$ which is a contradiction. Hence, it must be the case that  $(F_1, F_2)$ is  distinguishable, namely, $\overline{t}_1(Q_4) \geq 5.$

 \begin{theorem}
  \label{theorem:tQMM}
 $\overline{t}_1(Q_3, \mbox{ \rm MM*})=3,$  $\overline{t}_1(Q_4, \mbox{ \rm MM*})=5,$ and, for $n \geq 5$ and $g \in [1, n-3],$  $
\overline{t}_g(Q_n, \mbox{\rm MM*}) $ $=\frac{1}{2}(g+1)(2n-g). $
 \end{theorem}
 The above result as shown in Theorem~\ref{theorem:tQMM} agrees with the one achieved in\ \cite{Liu2017}, where  $n \geq 5,$ $g \in [1, \frac{n-1}{4}],$ and we notice that $n-3 > \frac{n-1}{4},$ when $n \geq 4.$  
 
 \smallskip
 
 We would like to point out that, the general $g$-extra connectivity result for the hypercube structure, as stated in Theorem~\ref{theorem:kQExtended}, can also be used to derive the following  $g$-extra diagnosability result for $Q_n$ under the MM* model, by applying the aforementioned general derivation process.
\begin{theorem}
\label{theorem:atQMM}
 $\overline{t}_1(Q_3, \mbox{ \rm MM*})=3,$  $\overline{t}_1(Q_4, \mbox{ \rm MM*})=5,$ and, for  $n \geq 5, $   $g \in [0, 3n-7],$    $\overline{t}_g(Q_n, \mbox{\rm MM*})=\overline{\kappa}_g(Q_n)+g,$ where $\overline{\kappa}_g(Q_n)$ is given in Theorem~\ref{theorem:kQExtended} with the proper and respective range of $g.$ 
 \end{theorem}
\noindent{\bf Proof:}  Beside the special cases when $n \leq 4,$ and $g=1,$ by Corollaries~\ref{corollary:kQ},~\ref{corollary:lubG}, the usual upper bound construction,  together with the justification as given in\ \cite[Lemma~3.1]{Zhu2016}, we have 
\[
\overline{t}_g(Q_n, \mbox{\rm MM*}) \leq \overline{\kappa}_g(Q_n)+g.
\]
 It is also a routine check to verify that, for all the cases, $|V(Q_n)|=2^n > 2(\overline{\kappa}_g(Q_n)+g).$ Hence, by Corollary~\ref{corollary:ge>=2}, for all $g \geq 2,$
 \[
\overline{t}_g(Q_n, \mbox{\rm MM*}) \geq \overline{\kappa}_g(Q_n)+g.
\]
This completes the proof of this result.

\section{The $g$-extra diagnosability of the $(n, k)$-star graph}
\label{section:resultNK}

The star graph, denoted by $S_n,$ was proposed in\ \cite{Akers1989} as an attractive alternative to the hypercube structure when used as an interconnection network. For comparison between the hypercube and the star graph, readers are referred to\ \cite{Day1994,Kiasari2008}. However, the requirement that the number of vertices in the star graph be $n!$ results in a large size gap between $S_n$ and $S_{n+1}.$ To address this scalability issue, the $(n, k)$-star graph was suggested in\,\cite{Chiang1995}, which brings in a flexibility in choosing its size, while preserving many attractive properties of the star graph, including vertex symmetry. The $(n, k)$-star graph has been well studied in the literature, including its fault-tolerant properties, e.g.,\ \cite{Yang2010,Yuan2011,CQS2012,Li2014,CQS2019,Lv2019}.

Let  $\langle n \rangle$ stand for $\{1, 2, \ldots, n\}$,  $V(S_{n, k}), n \geq 2, k \in [1, n),$ is simply the collection of all the $k$-permutations taken out of $\langle n \rangle,$ thus, $|V|=n!/(n-k)!.$ Let $u=[p_1, \cdots,  p_k],$ $v=[q_1, \cdots, q_k],$ $(p, q) \in E(S_{n, k})$ either, for some $i \in [2, k],$ $v$ can be obtained from $u$ by swapping $p_1$ and $p_i$\ ({\em $i$-edge}); or, for some $e\ (\in \langle n \rangle \setminus \{p_1, \cdots,  p_k\}),$ $v$ can be obtained from $u$ by replacing $p_1$ with $e$\ ({\em $1$-edge}). Thus, $S_{n,k}$ is an $n - 1$ regular graph, containing exactly $[(n-1)n!]/[2(n-k)!]$ edges.

It is easy to see and well known that the connectivity of the $(n, k)$-star graph is $n-1$\ \cite[Theorem~9]{Chiang1995}. Thus, by Corollary~\ref{corollary:first}, it is immediate that, for $g \geq 1,$ $t_g(S_{n, k}) \geq t_0(S_{n, k}) = n-1,$ and $\overline{t}_g(S_{n, k}) \geq \overline{t}_0(S_{n, k}) = n-1.$
\begin{figure}[bht]
\centering
\includegraphics[width=3in]{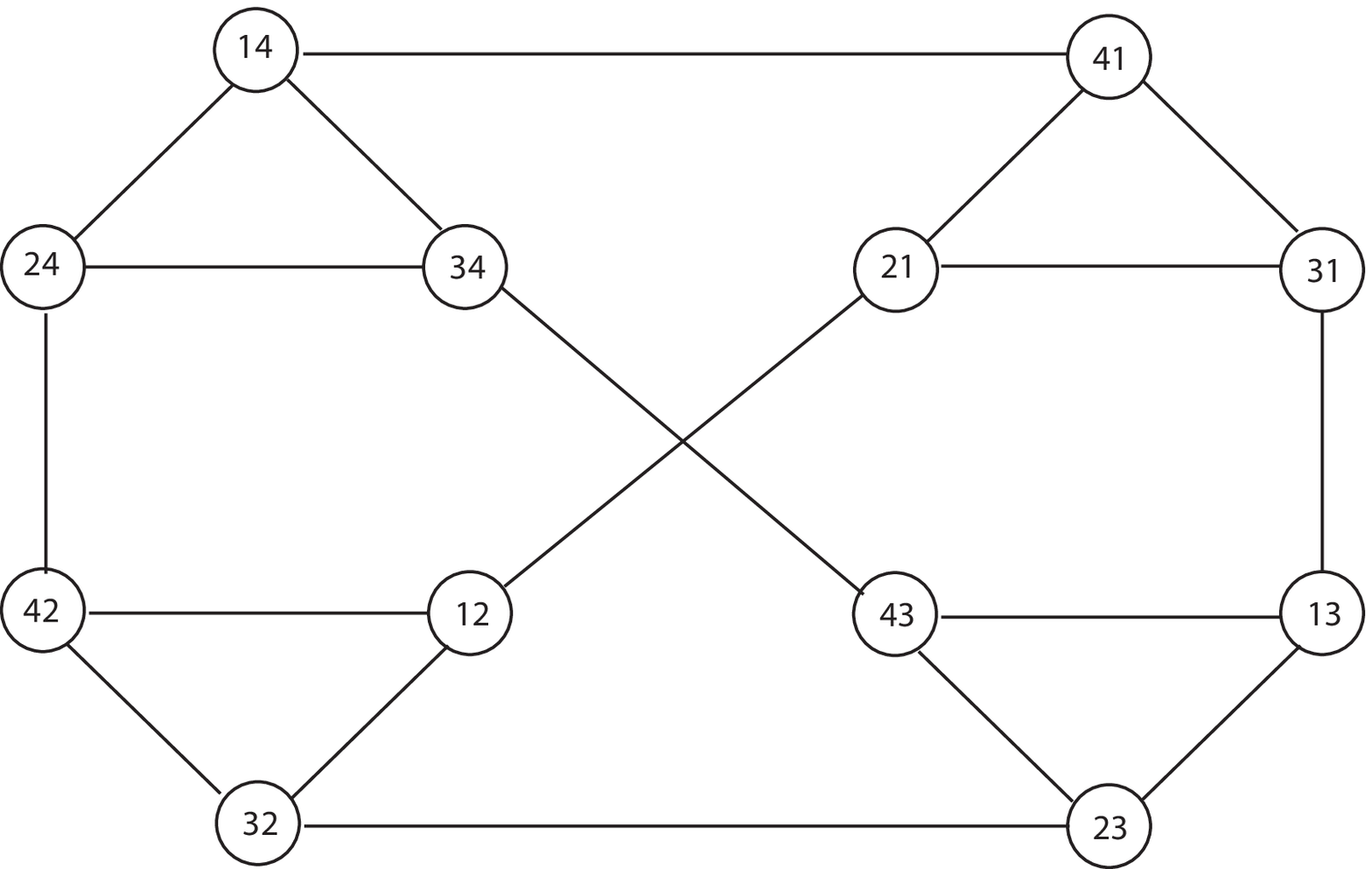}
\caption{$S_{4, 2}$}
\label{figure:s(42)}
\end{figure}

Let $H_i, i \in [1, n],$ be the collection of all the vertices of $S_{n, k},$ where the corresponding $k$-permutation ends with $p_k=i,$ and let $S_{n, k}^i$ be the sub-graph of $S_{n, k}$ with its vertex set being $H_i,$ it is easy to see and  well known that $S_{n, k}^i$ is isomorphic to $S_{n-1, k-1}.$ Moreover, every vertex in $S_{n, k}^i$ has a unique neighbor in $S_{n, k}^j, j \not = i,$ and for each pair of $S_{n, k}^i$ and $S_{n, k}^j,$ there are exactly $(n-2)!/(n-k)!$ independent edges connecting their respective vertices between. Readers are referred to\ \cite{Chiang1995,Yuan2011} for more details.  

For example,  $[1, 2]$\ (represented as $12$) is a vertex in  $S_{4, 2},$ as shown in Figure~\ref{figure:s(42)}, where  $([1, 2],[3, 2])$ and $([1, 2], [4, 2])$ are both $1$-edges, and $([1, 2], [2, 1])$ is a 2-edge. Clearly, $H_2=\{[1, 2], [3, 2], [4, 2]\},$ which are the vertices of $S_{4, 2}^2.$ Moreover, the vertex $[1, 2]$ has a  unique neighbor $[2, 1]$ in $S_{4, 2}^1,$ and this 2-edge $([1, 2], [2, 1])$ is a unique one between $S_{4, 2}^1$ and $S_{4, 2}^2.$

Issues related to the $g$-good-neighbor diagnosability of the $(n, k)$-star graph have been addressed in\ \cite{Xu2017,CQS2019}. Its $g$-extra diagnosability has also been derived recently in\ \cite{Lv2019} by following a structure dependent approach. It turns out that these results on $g$-extra diagnosability of the $(n, k)$-star
graph also follow the general result that we have derived in the previous section, thus all the structure independent technical details are unnecessary, and could be spared.

Since it has already been proved that $t_1(S_{n, k}, D)=n+k-2$\ (\ \cite[Theorem~5.3]{CQS2019}\ (PMC) and\ \cite{Xu2017}\ (MM*), by Corollary~\ref{corollary:g1=tg1}, we immediately have the following result.
\begin{corollary}
\label{corollary:tgN1}
Let $n \geq 4, k \in [2, n),$ $\overline{t}_1(S_{n, k}, D) = n+k-2.$
\end{corollary}

We now move forward to the cases of $g \geq 2.$ By Corollary~\ref{corollary:gnE} and Theorem~\ref{theorem:UBSg}, we obtain the following upper bound result for the $g$-extra diagnosability of the $(n, k)$-star graph.
\begin{lemma}
\label{lemma:upnk}
For $n \geq 4, k \in [2, n), g \in [0, n-k],$ $\overline{t}_g(S_{n, k}, D)\leq n+g(k-1)-1.$ 
\end{lemma}

We again make use of  the super-connectedness property of the $(n, k)$-star graph to derive the $g$-extra connectivity of various graphs\ \cite{Xu2016,Lv2019}, then its $g$-extra diagnosability. We start with the following observation.
\begin{theorem}
\label{theorem:nkg}
\cite[Theorem~8]{Yuan2011}
Let $n \geq 4, k,$ and $r$ be positive integers such that $k \in [2, n)$ and $ r \in [1, n-k+1].$ If $T$ is a set of vertices of $S_{n,k}$ such that $|T | \leq n+(r-1)k-2r,$ then $S_{n, k}-T$ is either connected or has a large component and small components with at most $r-1$ vertices in total.
\end{theorem}
Thus, for $r \in [1, n-k+1],$ $(n, k)$-star graphs are super $n+(r-1)k-2r$-vertex connected of order $r-1.$ Taking $r=g+1,$ if we want to have a component, beside the larger one, in a disconnected survival graph $S_{n, k}-T,$ which contains at least $g+1$ vertices, $g \in [0, n-k],$ we have to remove at least $n+g(k-2)-1$ vertices, i.e., 
\[
\overline{\kappa}_g(S_{n, k}) \geq n+g(k-2)-1.
\]

The above lower bound is actually tight. Indeed, as suggested in\ \cite{Yuan2011}, let $Y=\{u_1, u_2,\ldots, u_{g+1} \},$ where $u_j=[j, n-k+2, \ldots, n], j \in [1, g+1].$ $Y$ is well defined since $g+2 \leq n-k+2 \equiv g \leq n-k.$ By definition, each vertex $u_j$ has $k-1$  neighbors through $i$-edges, $i \in [2, k]$, thus $(g+1)(k-1)$ distinct $i$-neighbors in total, since each $u_j, j \in [1, g+1],$ starts with a distinct symbol. Moreover, since a total of $k+g$ symbols have occurred in $Y,$ there are $n-k-g$ distinct neighbors of vertices in $Y$ through 1-edges by switching $j \in [1, g+1]$ with $p \in [g+2, n-k+1].$ There are thus $n-k-g$ neighbors of all the vertices in $Y.$ As a result, $|N(Y)|=(g+1)(k-1)+(n-k-g)=n+g(k-2)-1.$

We notice that all these vertices in $Y$ belong to $S_{n, k}^n,$ and, all of the $n+g(k-2)-1$ neighbors as contained in $N(Y),$ except $g+1$ of them, also belong to $S_{n, k}^n.$ For $j \in [1, g+1],$ $u_j',$ each of these $g+1$ neighbors that do not belong to $S_{n k}^n,$   belongs to $S_{n, k}^j,$ respectively. It is clear that such a neighbor $u_j'\ (=[n, n-k+2, \ldots, j])$ is obtained from $u_j\ (=[j, n-k+2, \ldots, n])$ by swapping $j$ with $n.$   

Let $F=N(Y),$ and, for all $j \in [1, n],$ $F_j=H_j \cap F,$ we have that, for $j \in [1, g+1], |F_j|=1,$ for $j \in [g+2, n-1], |F_j|=0,$ and $|F_n|=n+g(k-2)-2.$ Thus, although $S_{n, k}^n-F$ is disconnected, for all $j \in [1, n-1],$ $S_{n, k}^j-F_j$ is connected. In particular, since, e.g., $|F_{g+2}|=0$, when $n > k \geq 3,$ for all $j \in [1, n-1]\setminus \{g+2\},$ there are at least $\frac{(n-2)!}{(n-k)!} \geq (n-2) \geq 2$ independent edges connecting $S_{n, k}^j$ and $S_{n, k}^{g+2}.$ Thus, for all $j \in [1, n-1],$ any $S_{n, k}^j-F_i$ belongs to the large component $Z_1.$  Notice that, when $n>k \geq 3 >2,$ $g \leq n-k < n-2,$ $|\{j, |F_j|=0\}|=n-g-2 \geq 1.$ Then
$|Z_1|\geq |H_{g+2}|=(n-1)!/(n-k)! \geq 3(n-1)> n-2 > g+1.$  

Moreover, each vertex $z$ in $S_{n, k}^n-F_n$ is adjacent to a unique neighbor in $S_{n, k}^j, j \in [1, g+1],$ thus not with any $u_j', j \in [1, g+1].$  Therefore, when we remove $N(Y)$ from $S_{n, k},$ the survival graph, $S_{n, k}-N(Y)$, contains a large component, $Z\ (=S_{n, k}-N^c(Y))$, $Z_1 \subseteq Z,$ and a small one, $Y,$ both containing at least $g+1$ vertices.   

Hence, $N(Y)$ is indeed a $g$-extra cut of $S_{n, k}$, and $\overline{\kappa}_g(S_{n, k}) \leq  n + g(k - 2) - 1.$ Combining with
the aforementioned lower bound result, we have the following $g$-extra connectivity result for $S_{n, k}.$
\begin{corollary}
\label{corollary:nkg}
Let $n \geq 4, k \in [3, n), g \in [1, n-k],$ $\overline{\kappa}_g(S_{n, k}) =n+g(k-2)-1.$
\end{corollary}

We are now ready to derive the following general result.
\begin{theorem}
\label{theorem:nkgP}
Let $n \geq 4, k \in [3, n), g \in [1, n-k].$ Then, $\overline{t}_g(S_{n, k}, \mbox{\rm PMC}) = n+g(k-1)-1.$
\end{theorem}
\noindent{\bf Proof:} Since $n \geq 4$, and $k \geq 3,$ $g \leq n-3,$ and $k \leq n-1,$ we have that
\begin{eqnarray*}
&&|V|-2(n+g(k-1)-1) \geq n(n-1)(n-2)-2[n+(n-4)(n-2)-1] \\
&=& n^3-5n^2+20n-14>0.
\end{eqnarray*}
The result now holds by Corollary~\ref{corollary:lbgPMC},~Corollary~\ref{corollary:nkg}, and Lemma~\ref{lemma:upnk}. 

\smallskip
We notice that the above result agrees with that obtained in\ \cite[Theorem~4.3]{Lv2019} with essentially the same ranges for $n, k$ and $g.$
\smallskip

We also have the following result by Corollary~\ref{corollary:ge>=2},~ Corollary~\ref{corollary:tgN1}, which provides the $g=1$ case, Corollary~\ref{corollary:nkg},~Lemma~\ref{lemma:upnk}, and the routine checking as made in the proof of the above result.
\begin{theorem}
\label{theorem:nkgMM}
Let $n \geq 4, k \in [3, n), g \in [1, n-k],$ $\overline{t}_g(S_{n, k}, \mbox{\rm MM*}) = n+g(k-1)-1.$
\end{theorem}
\smallskip
We notice that the above result also agrees with that obtained in\ \cite[Theorem~4.7]{Lv2019} , where $n \geq 6, k \in [3, n-3],$ and $g \in [1, \min\{k-2, \frac{n-k+1}{4} \}].$ 

It is worth pointing out that  Theorem~\ref{theorem:nkg} was recently restated in\ \cite[Lemma~3.3]{Lv2019} with a shorter proof, and an alternative, structure dependent, derivation was made in\ \cite{Lv2019} to obtain the $g$-extra diagnosability of the $S_{n, k}$ in terms of both the PMC and the MM* models, where much structure independent details, as we summarized in Sections~\ref{section:relation} and~\ref{section:process}, could be spared. 

It also holds that, since we have shown $N(Y)$ is a $g$-extra faulty set, and $|N^c(Y)|= n+g(k-1),$ we can also apply Corollary~\ref{corollary:lubMR} to obtain Lemma~\ref{lemma:upnk}. 

Incidentally, since $S_{n, n-1}$ is isomorphic to the star graph\ \cite[Lemma~4]{Chiang1995}, and $S_{n, n-2}$ is isomorphic to the alternating group network\ \cite{CQS2012b}, the $g$-extra diagnosability results of these latter two graphs immediately follow. 

Finally, we comment that, when taking $g = 0$ in both Theorems~\ref{theorem:nkgP} and~\ref{theorem:nkgMM}, we have that
\[
\overline{t}_0(S_{n, k}, \mbox{\rm PMC}) = \overline{t}_0(S_{n, k}, \mbox{\rm MM*}) = n - 1,
\]
i.e., the unrestricted diagnosability, or the vertex connectivity, of $S_{n, k},$ as expected.

\section{The $g$-extra diagnosability of the arrangement graph}
\label{section:resultA}

The  arrangement graph  is another alternative structure suggested in\ \cite{Day1992} to address the scalability issue as associated with the star graph \cite{Akers1989}.  This class of graphs also preserve many nice properties of the star graph such as vertex and edge symmetry, hierarchical and recursive structure, and simple shortest path routing. It has also drawn a considerable amount of attention with its various fault-tolerant properties
\ \cite{Hsieh1999,Zhou2011,Yuan2011,Cheng2013,CQS2013,Wang2014,CQS2019}. 

The vertex set of an arrangement graph, denoted by $A_{n, k}, n\geq 2, k \in [1, n),$ is also the collection of all the $k$-permutations taken out of $\langle n \rangle\ (=\{1, 2, \ldots, n\}),$ and two vertices are adjacent to each other if and only if they differ  in exactly one position. $A_{n, k}$ thus also contains exactly $n!/(n-k)!$ vertices. Let $u\ (=[p_1, p_2, \ldots, p_k])$ be a vertex of $A_{n, k},$ we can get $u',$ a neighbor of $u,$ by replacing $p_i, i \in [1, k],$ with any of the $n-k$ symbols that does not occur in $u.$ Thus, $A_{n, k}$ is a regular graph where the degree of all its vertices equals $k(n-k),$ which is also its connectivity\ \cite{Day1992}. 

Recall that $H_i, i \in [1, n],$ collects all the vertices where the corresponding $k$-permutation ends with $p_k=i,$ and let $A_{n, k}^i$ be the sub-graph of $A_{n, k}$ restricted on $H_i, i \in [1, n],$ it is also well known that $A_{n, k}^i$ is isomorphic to $A_{n-1, k-1}.$ Each vertex in $A_{n, k}^i$ is adjacent to exactly $n-k$ neighbors, one each in a different $A_{n, k}^j, j \not = i,$ and, for each pair of $A_{n, k}^i$ and $A_{n, k}^j,$ there are exactly $(n-2)!/(n-k-1)!$ independent edges connecting them. Readers are referred to\ \cite{Day1992,Cheng2013} for more detailed discussion of the structural properties of the arrangement graph. 

Let $x, y$ be two vertices in a graph $G(V, E),$ we use $d(x, y)$ to denote the distance of, i.e., the length of a shortest path between, $x$ and $y$ in $V$ in terms of $E.$  Recall that $N(u)$ stands for the neighbors of $u,$ we find the following common neighbor result useful. 
\begin{lemma}
\cite[Lemma~3]{Xu2016}
\label{lemma:neighbor}
Let $u, v$ be two vertices in $A_{n, k}, n\geq k-1,$ then
\[
|N(u) \cap N(v)|= \left \{
\begin{array}{ll}
0 & \mbox{ \rm if $d(u, v) \geq 3$}, \\
2 & \mbox{ \rm if $d(u, v)=2$ and $n \geq k+2$}, \\
1 & \mbox{ \rm if $d(u, v)=2$ and $n=k+1$ and}, \\
$n-k-1$ & \mbox{ \rm if $d(u, v)=1$}.
\end{array}
\right.
\]
\end{lemma}

\begin{figure}[hbt]
\centering
\includegraphics[width=2in]{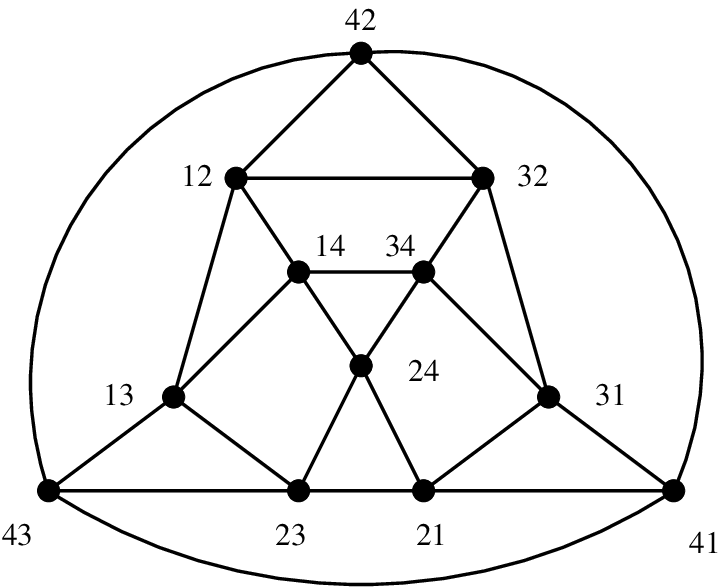}
\caption{ $A_{4, 2}$}
\label{figure:a(42)}
\end{figure}

Figure~\ref{figure:a(42)} shows $A_{4, 2},$ where, e.g., $[1, 2]$\ (represented as $12$), a vertex of $A_{4, 2}^2,$ has two neighbors, $[1, 3],$ a vertex of $A_{3, 2}^3,$ and $[1, 4],$ a vertex of $A_{4,2}^4.$ There are thus clearly two independent edges $\{([1, 2], [1, 3]), ([4, 2], [4, 3]) \}$  between $A_{4, 2}^2$ and $A_{4, 3}^3.$  
It is also clear that 12 and 13 share exactly one common neighbor, i.e., 14; 12 and 34 share two common neighbors, 14 and 32; while 12 and 31 share no neighbors, as mandated by Lemma~\ref{lemma:neighbor}.

Issues related to the $g$-good-neighbor diagnosability of the arrangement graphs have been addressed in\ \cite{Wang2018,CQS2019}. The $g$-extra diagnosability, $g \in \{1, 2, 3\},$ of the arrangement graph in terms of the PMC model has also been derived in\ \cite{Xu2016} by following a structure dependent approach. We will show that this diagnosability result, and that in terms of the MM* model, also naturally follow the general process that we have described in the previous section.  

Since it has been proved in\ \cite[Theorem~4.5]{CQS2019} that for $n \geq 5, k \in [2, n),$ $t_1(A_{n, k}, \mbox{\rm PMC})=(2k-1)(n-k),$ and\ \cite[Theorem~4.7]{CQS2019} that for $n \geq 6, k \in [5, n-1),$ or $n \geq 11, k \in [10, n),$ $t_1(A_{n, k}, \mbox{\rm MM*})=(2k-1)(n-k),$ the following results immediately follow
 Corollary~\ref{corollary:g1=tg1}.

\begin{theorem}
\label{theorem:tgA1PMC}
For $n \geq 5, k \in [2, n),$ $\overline{t}_1(A_{n, k}, \mbox{\rm PMC})=(2k-1)(n-k).$
\end{theorem}
We notice that the above result slightly generalizes the one obtained in\ \cite[Theorem~4]{Xu2016}, where $n \geq 5, k \in [3, n-2].$
\begin{theorem}
\label{theorem:tgA1MM}
For $n \geq 6, k \in [5, n-1),$ or $n \geq 11, k \in [10, n),$ $\overline{t}_1(A_{n, k}, \mbox{\rm MM*})=(2k-1)(n-k).$
\end{theorem} 
We now move to the case when $g=2,$ starting with the following super-connectedness property of the arrangement graphs.
\begin{theorem}
\cite{Cheng2013}
\label{theorem:gt=2}
Let $n \geq  8, k \in [2, n-5], $ and let $T$ be a subset of the vertices of $A_{n, k}$ such that $|T|\leq (3k-2)(n-k)-4.$ Then $A_{n, k}-T$ is either
connected or has a large component and small components with at most two vertices in total unless $k = 2$ and $|T|= 4n-12,$ in which case $A_{n, k}-T$ could have a large component and a 4-cycle.
\end{theorem} 

It is thus clear, by \cite[Theorem~4.2]{CHQS2019}, that if we want to end up with a survival graph, $A_{n, k}-T,$ where every component, beside the largest one, contains at least three vertices, $|T|$ has to be at least $(3k-2)(n-k)-3.$ It is also shown in\ \cite{Cheng2013} that such a bound is tight\ \footnote{The construction of such a tight $3$-extra cut will be given later.}. Hence, we have the following result.
\begin{corollary}
\label{corollary:Alb}
Let $n \geq 8, k \in [3, n-5], \overline{\kappa}_2(A_{n, k})=(3k-2)(n-k)-3.$
\end{corollary}

It is easy to see that, when $n \geq 4, k \in [3, n-5],$ 
\[
n!/(n-k)! \geq n(n-1)(n-2) \geq 2[(3(n-5)-2)(n-3)-1] >2[(3k-2)(n-k)-1].
\]
 Thus, by Corollaries~\ref{corollary:lbgPMC}\ (PMC), and~\ref{corollary:ge>=2}\ (MM*), we have achieved the following lower bound results for $2$-extra diagnosability for the arrangement graph.
\begin{corollary}
\label{corollary:2tgA}
Let $n \geq 8, k \in [3, n-5], \overline{t}_2(A_{n, k}, D) \geq (3k-2)(n-k)-1.$
\end{corollary}

To derive an upper bound for $\overline{t}_2(A_{n, k}),$ we notice that, by Corollary~\ref{corollary:ttg}, $\overline{t}_2(A_{n, k}, D) \leq {t}_2(A_{n, k}, D),$ where $D$ refers to either the PMC or the MM* model. On the other hand, we have achieved the following result earlier.
\begin{theorem}
\cite{CQS2019}
\label{theorem:Ngd=2}
For $n \geq 7, k \in [4, n -1),$ $t_2(A_{n, k}, D) =(3k -2)(n -k).$
\end{theorem}
Hence, we have the following result, which apparently is not a tight upper bound in light of Corollary~\ref{corollary:2tgA}.
\begin{corollary}
For $n \geq 7, k \in [4, n -1),$ $\overline{t}_2(A_{n, k}, D) \leq (3k -2)(n -k).$
\end{corollary}
Indeed, the $g$-extra fault-tolerant model is not as demanding as the $g$-good-neighbor fault-tolerant model, as characterized in Corollary~\ref{corollary:gnE}. We now follow the guidance of Proposition~\ref{proposition:lubM} to construct a tight upper bound. 

Consider the following length 2 path, $Y_1\ (=(u, v, w))$ in $A_{n, k}^k, k \in [4, n-2],$ where $u=[1, 2, 3, \ldots, k], v=[1, k+1, 3, \ldots, k],$ and $w=[k+2, k+1, 3, \ldots, k].$ Since there do not exist common neighbors of $u, v$ and $w$\ \footnote{ Since $d(u, w)=2,$ and $k \leq n-2,$ by Lemma~\ref{lemma:neighbor}, $u$ and $w$ share exactly two common neighbors, one of them being $v.$ Hence, $u, v$ and $w$ share no common neighbors, as $v$ cannot be its own neighbor. }, to identify $N(Y),$ out of $3k(n-k)$ neighbors of $u, v$ and $w,$ we need to 1) remove those in $Y$, 2) remove those neighbors shared by both $u$ and $v,$ 3) those shared by both $v$ and $w,$ and 4) those shared by $u$ and $w.$

Beside the fact that $Y_1=\{u, v, w\},$ $u$ and $w$ have exactly two neighbors: $v$ and $[k+2, 2, 3, \ldots, k]$, it is clear that, 
\begin{itemize}
  \item  $N(\{u, v\})=\{1, p_2, 3, \ldots, k \},$ where $p_2 \in [k+2, n].$ Thus, $|N(\{u, v\})|=n-k-1,$ consistent with Lemma~\ref{lemma:neighbor}. In other words, there are $n-k-1$ neighbors shared by both $u$ and $v.$ They are $[1, k+2, 3, \ldots, k], $ $[1, k+4, 3, \ldots, k],$ \ldots, and $[1, n, 3, \ldots, k],$ all falling into $A_{n, k}^k.$ 
  \item  $N(\{v, w\})=\{p_1, k+1, 3, \ldots, k \},$ where $p_1 \in \{2\} \cup [k+3, n].$ Thus, $|N(\{v, w\})|=n-k-1,$ also consistent with Lemma~\ref{lemma:neighbor}. Those neighbors  are $[2, k+1, 3, \ldots, k], $ $[k+3, k+1, 3, \ldots, k],$ \ldots, and $[n, k+1, 3, \ldots, k],$ also falling into $A_{n, k}^k.$ 
\end{itemize}  
We also notice that both $u$ and $w$ are neighbors of $v,$ each counted once as a neighbor of $v,$ and $v$ is a neighbor of both $u$ and $w$ in $Y_1,$ counted once as a neighbor of both $u$ and $w.$ Thus, by the Principle of Inclusion and Exclusion,
\begin{eqnarray}
|N(Y_1)| &=& |N(u)|+|N(v)|+N(w)| \nonumber \\
&&-[|N(u, v)+N(v, w)+N(u, w)]+|N(u, v, w)| \nonumber \\
&=&[3k(n-k)-4]-[2(n-k-1)+1] \nonumber \\
&=&(3k-2)(n-k)-3. \label{eq:g=2}
\end{eqnarray}
Thus, $|N^c(Y_1)|=|N^c(Y_1)|+|Y_1|=(3k-2)(n-k).$ 

\smallskip

We now proceed to show that $N(Y_1),$ clearly a vertex cut, is indeed a $2$-extra faulty set. In particular, we show that $A_{n, k}-N(Y_1)$ contains two components,  $Y_1$ and another, larger, component, referred to $Z_1$ in the later discussion, both containing at least 3 vertices. To this regard, we observe that, out of the $k(n-k)$ neighbors of $u,$ $\{[1, p_i, \ldots, k]: i \in [1, k], p_i \in [k+1, n]\},$  $(k-1)(n-k)$ of them, including $v\ (=[1, k+1, 3, \ldots, k]),$ fall into $A_{n, k}^k,$ when taking $i \in [1, k-1],$ and the other $n-k$ of them, $\{[1, 2, \ldots, p_k]: p_k \in [k+1, n]\},$ fall into $A_{n, k}^j, j \in [k+1, n].$ 

Moreover, out of the $k(n-k)$ neighbors of $v,$ $\{[1, k+1, p_i, \ldots, k]: i \in [1, k], p_i \in \{2\} \cup [k+2, n]\},$  $(k-1)(n-k)$ of them, including both $u\ (=[1, 2, \ldots, k])$ and $v\ (=[1, k+1, 3, \ldots, k]),$ fall into $A_{n, k}^k,$ when taking $i \in [1, k-1],$ and the other $n-k$ of them, $\{[1, k+1, , \ldots, p_k]: p_k \in \{2\}\cup [k+2, n]\},$ fall into $A_{n, k}^j, j \in \{2\} \cup [k+2, n].$ 

Finally. out of the $k(n-k)$ neighbors of $w,$ $\{[k+2, k+1, p_i, \ldots, k]: i \in [1, k], p_i \in \{1, 2\} \cup [k+3, n]\},$  $(k-1)(n-k)$ of them, including  $v\ (=[k+2, k+1, 3, \ldots, k]),$ fall into $A_{n, k}^k,$ when taking $i \in [1, k-1],$ and the other $n-k$ of them, $\{[k+2, k+1, , \ldots, p_k]: p_k \in \{1, 2\}\cup [k+3, n]\},$ fall into $A_{n, k}^j, j \in \{1, 2\} \cup [k+3, n].$ 

As discussed earlier, we also know that each of $N(\{u, v\})$ and $N(\{v, w\})$ contains $n-k-1$ vertices, and $N(\{u, w\})$ contains one vertex, all falling into $A_{n, k}^k.$

To summarize, let $F=N(Y_1),$ and for all $i \in [1, n],$ let $F_i=H_i \cap F,$ we have that for all $i \in [3, k-1], |F_i|=0,$ $|F_1|=|F_{k+1}|=1,$ $|F_2|=|F_{k+2}|=2,$ for all $i \in  [k+3, n], |F_i|=3,$  and $|F_k|=(3k-5)(n-k)-3$\ \footnote{We comment that $\sum_{i=1}^n |F_i|=(3k-2)(n-k)-3,$ as expected.}. Thus,  for all $i \in [1, n], i \not = k,$ $|F_i| \leq 3,$ it follows that, for all $i, j \in [1, n]\setminus \{k\},$ $|F_i|+|F_j| \leq 6.$

Assuming $4 \leq k \leq n-2,$ since there are $(n-2)!/(n-k-1)!\geq (n-2)(n-3)(n-4) \geq 24>6$\ \footnote{We notice that, if we require $k \in [3, n-2],$ we would have $n \geq k+1 \geq 4,$ and there would have $(n-2)(n-3)=6$ independent edges between. On the other hand, if $k \in [4, n-1],$ then $n \geq k+1=5,$ and there would be $(n-2)(n-3)(n-4)$ independent edges between, which also leads to six. This is why we have to require $k \in [4, n-2].$} independent edges between  $A_{n, k}^i$ and $A_{n, k}^j, k \not \in \{i, j\},$  each such a $A_{n, k}^i-F_i$ is a part of one large component, $Z_1,$ via such independent edges.  Moreover, let $u \in A_{n, k}^k - F_k,$ it has $n-k\geq 2$ unique outside neighbors, none being a neighbor of a vertex in $Y_1\ \subset H_k.$ Hence, all vertices in $A_{n, k}^k-F$ belong to this large component $Z_1,$ as well. 

It is clear that $|Y_1|=3,$ and, by assumption, for at least one $i \in [3, k-1], |F_i|=0,$ thus, $|V(H_i)|=(n-1)!/(n-k)!\geq (n-1)(n-2)(n-3) \geq 60>3$. It is thus also clear that $Z_1,$ containing at least one such  $A_{n, k}^i,$ contains more than three vertices. Therefore, $N(Y_1)$ is indeed a 2-extra faulty set of $A_{n, k}, k \in [4, n-2].$
Hence, $\overline{\kappa}_2(A_{n, k}) \leq |N(Y_1)|=(3k-2)(n-k)-3,$ verifying Corollary~\ref{corollary:Alb}. 

\smallskip
Finally, when $k \in [4, n-2],$ 
\begin{eqnarray*}
|V(A_{n, k})|&=&n!/(n-k)! \geq n(n-1)(n-2)(n-3)>[3(n-2)-2](n-4)\\
&\geq& (3k-2)(n-k)=|N^c(Y_1)|. 
\end{eqnarray*}
By Corollary~\ref{corollary:lubMR}, we have the following result.

\begin{corollary}
\label{corollary:2ubA}
For $n \geq 6, k \in [4, n-2],$ $\overline{t}_2(A_{n, k}, D) \leq (3k-2)(n-k)-1.$
\end{corollary}

Combining Corollaries~\ref{corollary:2tgA} and~\ref{corollary:2ubA}, we have the following tight bound result of the $2$-extra diagnosability of the arrangement graphs.
\begin{theorem}
\label{theorem:2gtA}
Let $n \geq 8, k \in [3, n-5],$ $\overline{t}_2(A_{n, k}, D)=(3k-2)(n-k)-1.$
\end{theorem}
The above result, when $D$ refers to the PMC model, agrees with that as shown in\ \cite[Theorem~5]{Xu2017}, where $n \geq 6, k \in [4, n-2].$

We comment that, if we use a 3-cycle, $Y_1'=(u, v, w),$ instead of the length 2 path $Y_1,$ where $u=[1, 2, \ldots, k],$ $v=[k+1, 2, \ldots, k],$ and $w=[k+2, 2, \ldots, k]$, let
\begin{eqnarray*}
C &=& \{[k+3, 2, \ldots, k], \ldots, [n, 2, \ldots, k]  \}, \\
D &=& \{[1, p_i, \ldots, k]:i \in [2, k], p_i \in [k+1, n]\}, \\
E &=& \{[k+1, p_i, \ldots, k]: i \in [2, k], p_i \in \{1\} \cup [k+2, n]\}, \mbox{ \rm and}\\
F &=& \{[k+2, p_i, \ldots, k]:i \in [2, k], p_i \in \{1, k+1\} \cup [k+3, n]\},
\end{eqnarray*}
we would have $N(u)= C \cup D, N(v) =  C \cup E, N(w) =  C \cup F,$ and $N(u, v)=N(v, w)=N(u, w)=N(u, v, w) =C.$

It is clear that $|C|=n-k-2,$ and $|D|=|E|=|F|=(k-1)(n-k).$ By the Principle of Inclusion-Exclusion, we get the following:
\begin{eqnarray*}
|N(Y_1')| &=& |N(u)|+|N(v)|+|N(w)| \\
&& -[|N(u, v)|+|N(v, w)|+|N(u, w)|]+|N(u, v, w)| \\
&=& 3|C|+(|D|+|E|+|F|)-3|C|+|C|\\
&=& |C|+|D|+|E|+|F| = (n-k-2)+3(k-1)(n-k) \\
&=& (3k-2)(n-k)-2 > |N(Y_1)|.
\end{eqnarray*}
Thus, for the case of $g=2$, a length 2 path provides a smaller upper bound of $g$-extra diagnosability as compared with a 3-cycle, in light of Proposition~\ref{proposition:lubM}. 
\smallskip

When moving towards the case of $g=3,$ we notice the following $g$-extra connectivity result appears in\ \cite[Lemma~6]{Xu2016}, which appeared earlier in\ \cite{Lin2014}.
\begin{theorem}
\cite{Xu2016}
For $n \geq 6, k \in [3, n-3]$ or $k \in [4, n-2],$ $\overline{\kappa}_3(A_{n, k})=4(k-1)(n-k)-4.$ 
\end{theorem}

Since, when $n \geq 6, k \in [3, n-3],$ we have 
\[
n!/(n-k)! \geq n(n-1)(n-2) \geq 2(4(n-4)(n-3)-1)+4 \geq 2(4(k-1)(n-k)-1)+4,
\] 
by Corollary~\ref{corollary:lbgPMC}\ (PMC), and Corollary~\ref{corollary:ge>=2}\ (MM*), we immediately have the following lower bound results for the $3$-extra diagnosability.
\begin{corollary}
\label{corollary:3LUBtgA}
Let $n \geq 6, k \in [3, n-3], \overline{t}_3(A_{n, k}, D) \geq 4(k-1)(n-k)-1.$
\end{corollary}
 
To the best of our knowledge, no results regarding $t_3(A_{n, k})$ exist. Thus, we cannot use Corollary~\ref{corollary:gnE} to get even an estimate of the upper bound of $\overline{t}_3({A_{n, k}, D}).$ We now follow the construction as discussed in Section~\ref{section:UB} to seek such an upper bound, making use of an example originally suggested in proving\ \cite[Theorem~3]{Xu2016}. 

Considering the following four-cycle, $Y_2\ (=(u, v, w, x))$ in $A_{n, k}^k, n \geq 7, k \in [4, n-2],$ where $u=[1, 2, 3, \ldots, k],$ $v=[k+1, 2, 3, \ldots, k],$ $w=[k+1, k+2, 3, \ldots, k]$ and $x=[1, k+2, 3, \ldots, k].$ Again, by Lemma~\ref{lemma:neighbor},  no vertex could be a neighbor of all the three vertices in $Y_1.$ To identify $N(Y_2),$ out of $4k(n-k)$ neighbors of $u, v$ and $w,$ we also need to 1) remove those in $Y_2$, and 2) remove those neighbors shared by both $u$ and $v,$ $v$ and $w,$ $u$ and $w,$ and also by $w$ and $x,$

Similar to the analysis made to derive $N(Y_1),$ we can find out that $|N(u, v)|=|N(v, w)|=N(w, x)|=|N(x, u)|=n-k-1.$ Considering that each of the four vertices in $Y_2$ is a neighbor of two other vertices in $Y_2,$ we have that, again by the Principle of Inclusion-Exclusion,
\[
|N(Y_2)|=4k(n-k)-8-4(n-k-1)=4[(k-1)(n-k)-1]. 
\]
Thus, $|N^c(Y_2)|=4(k-1)(n-k).$

We comment that, if we use a length 3 path, $Y_2'=(u, v, w, x),$ in the construction, since the distance between $u$ and $x$ is 3, none of their neighbors could be shared by Lemma~\ref{lemma:neighbor}. As a result, for $k \leq n,$ 
\[
|N(Y_2')|= (4k-3)(n-k)-3 > 4(k-1)(n-k)-4=|N(Y_2)|.
\]
Thus, for the case of $g=3,$ a 4 cycle is a better choice as compared with a length 3 path. 

\smallskip

We proceed to show that $A_{n, k}-N(Y_2)$ contains two components, a large component  $Z_2$, and $Y_2,$ both containing at least 4 vertices. 

Let $F=N(Y_2),$ and for all $i \in [1, n],$ let $F_i=H_i \cap F_1,$ we have also found, through an analysis similar to the case of $g=2,$ that, for all $i \in [3, k), |F_i|=0,$ $|F_1|=|F_2|=|F_{k+1}|=|F_{k+2}|=2,$ $|F_k|=4(k-2)(n-k)-4,$ and, for all $i \in [k+3, n],$ $|F_i|=4.$\ \footnote{It is based on this analysis that we set  $k \geq 4,$ hence $n \geq 7.$} Thus,  for all $i \in [1, n], i \not = k,$ $|F_i| \leq 4,$ it  follows that, for all $i, j \in [1, n]\setminus \{k\},$ $|F_i|+|F_j| \leq 8.$

Again, assuming that $4 \leq k \leq n-2,$ since $n \geq 7,$ there are $(n-2)!/(n-k-1)!\geq (n-2)(n-3)(n-4) \geq 60>8$ independent edges between  $A_{n, k}^i$ and $A_{n, k}^j, k \not \in \{i, j\},$  all such $A_{n, k}^i-F_i$ are connected into one large component $Z_2'$, via such independent edges.  Moreover, let $u \in A_{n, k}^k - F_k,$ it has $n-k \geq 2$ unique outside neighbors, none being a neighbor of a vertex in $Y_2\ \subset V(H_k).$ Hence, all vertices in $A_{n, k}^k-F$ belong to the same component containing $Z_2',$ forming a large component $Z_2.$  

It is clear that $|Y_2|=4,$ and, by assumption, for  $i \in [3, k], |F_i|=0,$ thus, $|V(H_i)|=(n-1)!/(n-k)!\geq (n-1)(n-2)(n-3) \geq 120>4,$ as $n \geq 7.$ It is thus also clear that $Z_2,$ which contains at least two such  $A_{n, k}^i$'s, also contains more than four vertices. Therefore, $N(Y_2)$ is indeed a 3-extra faulty set of $A_{n, k}, k \in [4, n-2].$  Finally, when $k \in [4, n-2],$ 
\begin{eqnarray*}
|V(A_{n, k})|&=&n!/(n-k)! \geq n(n-1)(n-2)(n-3)>4(n-3)(n-4)\\
&\geq& 4(k-1)(n-k)>|N^c(Y_2)|. 
\end{eqnarray*}
 Hence, by Corollary~\ref{corollary:lubMR}, 
\[
\overline{t}_3(A_{n, k}, D) \leq |N^c(Y_2)|-1=|N(Y_2)|+|Y_2|-1=4(k-1)(n-k)-1.
\] 
Together with Corollary~\ref{corollary:3LUBtgA}, we have the following result.
\begin{theorem}
\label{theorem:3tgA}
Let $n \geq 7, k \in [4, n-3], \overline{t}_3(A_{n, k}, D) = 4(k-1)(n-k)-1.$
\end{theorem}
The PMC version of the above result agrees, with a slightly smaller range, with that as obtained in\ \cite[Theorem~6]{Xu2017}, where $n \geq 6, k \in [3, n-3].$

It is well known that $A_{n, 1}$ is isomorphic to $K_n,$ the complete graph with $n$ vertices. We notice that both the $g$-good-neighbor and $g$-extra diagnosability of $K_n$, thus $A_{n, 1},$ have been derived in\ \cite[Theorem~11]{Wang2016c}  in terms of the PMC and the MM* model. We also notice that, since $A_{n, n-1}$ is isomorphic to the star graph\ \cite[Lemma~4]{Chiang1995}, and $A_{n, n-2}$ is isomorphic to the alternating group graph\ \cite{Jwo1993}, the $g$-extra diagnosability results of these latter two graphs immediately follow. For example, Theorems~\ref{theorem:tgA1PMC}, and~\ref{theorem:tgA1MM} do agree with Corollary~\ref{corollary:tgN1}, when taking $k=n-1.$ 

\smallskip

Incidentally, by Corollary~\ref{corollary:gnE}, Theorems~\ref{theorem:UBAg} and~\ref{theorem:3tgA}, we have the following range for $t_3(A_{n, k}),$ the 3-good-neighbor diagnosability of $A_{n, k}.$ 
\begin{corollary}
For $n \geq 6, k \in [3, n-3],$
\[
4(k-1)(n-k)-1 \leq t_3(A_{n, k}, D) \leq 4(k-1)(n-k)+n-k.
\]
\end{corollary}

\section{Concluding remarks}

\label{section:end}

In this paper, we explored general relationships among various  ``fault-free'' fault-tolerant models, where only fault-free vertices are to satisfy the required properties, and discussed the connection between such  fault-tolerant models,  and the diagnosability notion, consistent with the ``maximum restriction'' requirement. We then generalized a uniform  process that we can effectively apply to derive diagnosability results of various interconnection networks under the $g$-good-neighbor model and the $g$-extra fault-tolerant models, in terms of mainstream diagnostic models such as the PMC and the MM* models.  

As demonstrating examples, we showed how to apply such a general process to obtain $g$-extra diagnosability results for the hypercube, the $(n, k)$-star graph,  and the arrangement graph.  These results agree with those achieved individually, without duplicating structure independent technical details. Some of these results come with a larger range of application, and the result for the arrangement graph, for $g=3,$ in terms of the MM* model is new.  It is clear that such a general process can be applied to other interconnection networks to obtain their diagnosability results, assuming the associated connectivity result of such a graph is available, and appropriate construction can be identified for the fault-tolerant models. 

As future research topics,  beside studying other interconnection structures under those existing fault-tolerant models in light of this general process, we would also look into other appropriate fault-tolerant models, and the feasibility of applying this general process to derive various fault-tolerant properties under such alternative models. 

Beside the fact that the lower bound of an $M$-diagnosability result directly depends on the connectivity property related to the fault-tolerant model $M,$  the upper bound part, i.e., the construction of a pair of appropriate indistinguishable faulty sets, certainly depends on such a model, as well. Although the upper bound construction fits well with both the $g$-good-neighbor and the $g$-extra fault-tolerant models, it might not with other models. Beside the conditional fault-tolerant model related example as we gave in Section~\ref{section:UB}, as another example,  a set of vertices $F$ in a connected non-complete graph $G$ is called a {\em cyclic vertex-cut} if the survival graph $G -F$ is disconnected and at least two components in the survival graph contain a cycle\ \cite{CHQS2019}. It is straightforward to come up with a notion of  a $g$-cyclic faulty set when at least two components contain a cycle of length at least $g\ (\geq 3).$ Various related cyclic connectivity results have been achieved for the hypercube, the star graph, and other Cayley graphs generated by a transposition tree\ \cite[Chapter\,4]{CHQS2019}. On the other hand, it is clear that the $(N(Y), N^c(Y))$ construction that we used in this paper does not  fit in the context of this alternative fault-tolerant model since although, taking $Y$ as a $g$-cycle, $N(Y)$ could be a $g$-cyclic faulty set, i.e., it consists of at least two components, each containing a cycle, $N^c(Y)$ might not be, since it includes $Y,$ where a cycle resides. 

Therefore,  to derive the $M$-diagnosability of an interconnection network $G$   for a given fault-tolerant model $M$ and a diagnostic model $D,$ we need to derive the  $M$ connectivity of  $G$ to get the lower bound of such a diagnosability, and choose a pair of appropriate indistinguishable $M$-faulty sets for $G$ in terms of $D$ to establish its upper bound.

\end{document}